# Computational Optimal Transport: Complexity by Accelerated Gradient Descent Is Better Than by Sinkhorn's Algorithm

**Pavel Dvurechensky** [1]  **Alexander Gasnikov** [2 3 4]  **Alexey Kroshnin** [2 3 4]

## Abstract

We analyze two algorithms for approximating the general optimal transport (OT) distance between two discrete distributions of size $n$, up to accuracy $\varepsilon$. For the first algorithm, which is based on the celebrated Sinkhorn's algorithm, we prove the complexity bound $\widetilde{O}\left(\frac{n^2}{\varepsilon^2}\right)$ arithmetic operations [1]. For the second one, which is based on our novel Adaptive Primal-Dual Accelerated Gradient Descent (APDAGD) algorithm, we prove the complexity bound $\widetilde{O}\left(\min\left\{\frac{n^{9/4}}{\varepsilon}, \frac{n^2}{\varepsilon^2}\right\}\right)$ arithmetic operations. Both bounds have better dependence on $\varepsilon$ than the state-of-the-art result given by $\widetilde{O}\left(\frac{n^2}{\varepsilon^3}\right)$. Our second algorithm not only has better dependence on $\varepsilon$ in the complexity bound, but also is not specific to entropic regularization and can solve the OT problem with different regularizers.

## 1. Introduction

Optimal transport (OT) distances between probability measures or histograms, including the earth mover's distance (Werman et al., 1985; Rubner et al., 2000) and Monge-Kantorovich or Wasserstein distance (Villani, 2008), play an increasing role in different machine learning tasks, such as unsupervised learning (Arjovsky et al., 2017; Bigot et al., 2017), semi-supervised learning (Solomon et al., 2014), clustering (Ho et al., 2017), text classification (Kusner et al., 2015), as well as in image retrieval, clustering and classification (Rubner et al., 2000; Cuturi, 2013; Sandler & Lindenbaum, 2011), statistics (Ebert et al., 2017; Panaretos & Zemel, 2016), and other applications (Kolouri et al., 2017).

Our focus in this paper is on the computational aspects of OT distances for the case of two discrete probability measures with support of equal[2] size $n$. The state-of-the-art approach (Cuturi, 2013) for this setting is to apply Sinkhorn's algorithm to the entropy-regularized OT optimization problem. As it was recently shown in (Altschuler et al., 2017), this approach allows to find an $\varepsilon$-approximation for an OT distance in $\widetilde{O}\left(\frac{n^2}{\varepsilon^3}\right)$ arithmetic operations. In terms of the dependence on $n$, this result improves on the complexity $\widetilde{O}(n^3)$ achieved by the network simplex method or interior point methods (Pele & Werman, 2009), applied directly to the OT optimization problem, which is a linear program (Kantorovich, 1942). Nevertheless, the cubic dependence on $\varepsilon$ prevents approximating OT distances with good accuracy.

On the other hand, in image color transfer (Pitié et al., 2007) or domain adaptation (Courty et al., 2017) not only the OT distance, but also the optimal transportation plan is of interest. Recent works (Essid & Solomon, 2017; Blondel et al., 2017) observe that entropic regularization of the OT problem leads to a dense transportation plan, which is in contrast to the sparse transportation plan obtained by solving the unregularized OT problem. Motivated by this observation, they study general regularization by a strongly convex function, e.g. squared euclidean norm, and show that this leads to a sparse transportation plan. In this situation, Sinkhorn's algorithm becomes inapplicable since it is specific to entropic regularization.

Our goal in this paper is, first, to obtain better than state-of-the-art complexity bounds for approximating the OT distance and, second, propose a flexible algorithm for solving the OT problem with different types of regularization.

Approximating the OT distance amounts to solving the *OT*

---

[1]Weierstrass Institute for Applied Analysis and Stochastics, Berlin, Germany [2]National Research University Higher School of Economics, Moscow, Russian Federation [3]Moscow Institute of Physics and Technology, Dolgoprudny, Moscow Region, Russia [4]Institute for Information Transmission Problems RAS, Moscow, Russia. Correspondence to: Pavel Dvurechensky <pavel.dvurechensky@wias-berlin.de>.

[1]$\widetilde{O}$ hides polylogarithmic factors $(\ln n)^c$, $c > 0$.

[2]This is done for simplicity and all the results easily generalize to the case of measures with different support size.

*problem* (Kantorovich, 1942):

$$\min_{X \in \mathcal{U}(r,c)} \langle C, X \rangle,$$

$$\mathcal{U}(r,c) := \{X \in \mathbb{R}_+^{n \times n} : X\mathbf{1} = r, X^T\mathbf{1} = c\}, \quad (1)$$

where $X$ is *transportation plan*, $C \in \mathbb{R}_+^{n \times n}$ is a given ground cost matrix, $r, c \in \mathbb{R}^n$ are given vectors from the probability simplex $\Delta^n$, $\mathbf{1}$ is the vector of all ones. The *regularized OT problem* is

$$\min_{X \in \mathcal{U}(r,c)} \langle C, X \rangle + \gamma \mathcal{R}(X), \quad (2)$$

where $\gamma > 0$ is the *regularization parameter* and $\mathcal{R}(X)$ is a strongly convex *regularizer*, e.g. negative entropy or squared Euclidean norm.

Our goal is to find $\widehat{X} \in \mathcal{U}(r,c)$ such that

$$\langle C, \widehat{X} \rangle \leq \min_{X \in \mathcal{U}(r,c)} \langle C, X \rangle + \varepsilon. \quad (3)$$

In this case, $\langle C, \widehat{X} \rangle$ is an $\varepsilon$-approximation for the OT distance and $\widehat{X}$ is an approximation for the transportation plan.

**Related work.** We focus on the general case with $C$ being a non-negative dense matrix. In this case, (1) is a linear programming problem with best theoretical complexity $\widetilde{O}(n^{5/2})$ (Lee & Sidford, 2014) and best practical complexity $\widetilde{O}(n^3)$ (Pele & Werman, 2009), which is problematic when $n$ is larger than $10^3$.

A natural alternative is to approximate (1) by (2) with a small $\gamma$. Starting with the work (Cuturi, 2013), the widely used practical implementation of this idea is to use entropy regularization, i.e. solve (2), where $\mathcal{R}(X)$ is negative entropy of a matrix $X$. The special structure of this problem allows to use the balancing algorithm (Bregman, 1967) also known as Sinkhorn's algorithm (Sinkhorn, 1974) and RAS (Kalantari & Khachiyan, 1993). The best known complexity bound in the literature for this approach is $\widetilde{O}\left(\frac{n^2}{\varepsilon^3}\right)$ to obtain (3) (Altschuler et al., 2017), Theorem 1. They also show that the regularization parameter should be chosen proportional to $\varepsilon$, which necessitates working with the matrix $\exp(-C/\varepsilon)$ and leads to problems with numerical stability of the algorithm. Several ways to overcome this instability issue were proposed in (Schmitzer, 2016), but with limited theoretical analysis. While the entropy-regularized OT problem allows to use other matrix-scaling algorithms such as (Allen-Zhu et al., 2017; Cohen et al., 2017) with theoretical guarantees, the authors do not provide any experimental results, so the practical implementability of these algorithms is questionable. In (Genevay et al., 2016), stochastic gradient descent is applied to solve the entropy-regularized OT problem, but the complexity for approximating OT distance in the sense of (3) is not studied. In any case, Sinkhorn's

*Table 1.* Comparison of algorithms for (2).

| Algorithm | Rates | LS | Entr. |
|---|---|---|---|
| (Beck & Teboulle, 2014) | × | √ | √ |
| (Chambolle & Pock, 2011) | × | × | × |
| (Malitsky & Pock, 2016) | × | √ | × |
| (Tran-Dinh & Cevher, 2014) | √ | × | √ |
| (Yurtsever et al., 2015) | √ | ×[3] | √ |
| (Patrascu et al., 2015) | √ | × | √ |
| (Gasnikov et al., 2016) | √ | × | √ |
| (Chernov et al., 2016) | √ | × | √ |
| (Anikin et al., 2017) | √ | × | √ |
| (Dvurechensky et al., 2016) | × | √ | √ |
| (Bogolubsky et al., 2016) | × | √ | × |
| (Li et al., 2016) | √ | × | √ |
| (Lan et al., 2011) | × | × | √ |
| (Ouyang et al., 2015) | × | √ | √ |
| (Xu, 2016) | √ | × | × |
| This paper (Alg. 3) | √ | √ | √ |

and other mentioned algorithms are very specific to entropic regularization in (2).

A flexible alternative can be to use some general purpose optimization method to solve (2), which is a particular case of a minimization problem with linear constraints. When $n$ is large, the natural choice is the class of first-order methods, e.g. Conjugate Gradients (CG), quasi-Newton methods like L-BFGS or Nesterov's accelerated gradient descent (AGD). Due to the presence of linear constraints, the most common approach involves the construction of the Lagrange dual problem, which is an unconstrained problem. Based on the latter fact, L-BFGS was used in (Cuturi & Peyré, 2016; Blondel et al., 2017) for the dual problem. Since our focus is on complexity analysis, it is crucial to estimate the rate of convergence for the norm of the dual problem objective gradient as this norm is exactly the equality constraints feasibility in the primal problem. This can be complicated if CG or L-BFGS is used for the dual problem, so we choose AGD-type methods with primal-dual updates. The tricky part of this approach is to prove accelerated (Nesterov, 2004) convergence rates separately for the primal objective residual and linear constraints feasibility. On the other hand, first-order methods use the Lipschitz constant of the objective's gradient to define the stepsize. The theoretical value for this constant is usually an overestimation and leads to small stepsize and slow convergence in practice. Thus, an algorithm should use a line-search strategy to adapt to the local value of this constant and converge faster. Finally, entropy regularization in (2) is an important particular case and an algorithm should be able to deal with this non-Lipschitz-smooth regularizer. We analyzed a bunch of algorithms in the literature (see Table 1) and none of them combine all three described features, namely, *a) accelerated convergence rates separately for the primal objective and constraints feasibility, b) line-search, c) entropy friendliness.*

[3] Their algorithm uses Lipschitz constant in the stopping crite-

**Our contributions** can be summarized as follows.

- Improved analysis of the Sinkhorn's algorithm and complexity $\widetilde{O}\left(\frac{n^2}{\varepsilon^2}\right)$ arithmetic operations for approximating the OT distance in the sense of (3).

- An Adaptive Primal-Dual Accelerated Gradient Descent (APDAGD) algorithm, which incorporates a line-search strategy and has accelerated convergence rates separately for the primal objective and constraints[4] feasibility in (2) with a general strongly convex regularizer.

- Improved complexity $\widetilde{O}\left(\min\left\{\frac{n^{9/4}}{\varepsilon}, \frac{n^2}{\varepsilon^2}\right\}\right)$ arithmetic operations for approximating the OT distance in the sense of (3), based on our APDAGD method.

- Numerical illustration of the practical performance of these algorithms for approximating the OT distance.

**Notation.** For a general finite-dimensional real vector space $E$, we denote by $E^*$ its dual, given by linear pairing $\langle g, x \rangle$, $x \in E$, $g \in E^*$; by $\|\cdot\|_E$ the norm in $E$ and by $\|\cdot\|_{E,*}$ the norm in $E^*$, which is dual to $\|\cdot\|_E$. For a linear operator $A : E \to H$, we define its norm as $\|A\|_{E \to H} = \max_{x \in E, u \in H^*}\{\langle u, Ax \rangle : \|x\|_E = 1, \|u\|_{H,*} = 1\}$. We say that a function $f : E \to \mathbb{R}$ is $\gamma$-strongly convex on a set $Q \subseteq E$ w.r.t. a norm in $E$ iff, for any $x, y \in Q$, $f(y) \geq f(x) + \langle \nabla f(x), y - x \rangle + \frac{\gamma}{2}\|x - y\|_E^2$, where $\nabla f(x)$ is any subgradient of $f(x)$ at $x$.

For a matrix $A$ and a vector $a$, we denote by $e^A$, $e^a$, $\ln A$, $\ln a$ their entrywise exponents and natural logarithms respectively. For a vector $a \in \mathbb{R}^n$, we denote by $\|a\|_1$ the sum of absolute values of its elements, and by $\|a\|_2$ its Euclidean norm, and by $\|a\|_\infty$ the maximum absolute value of its elements. Given a matrix $A \in \mathbb{R}^{n \times n}$, we denote by $\text{vec}(A)$ the vector in $\mathbb{R}^{n^2}$, which is obtained from $A$ by writing its columns one below another. For a matrix $A \in \mathbb{R}^{n \times n}$, we denote $\|A\|_1 = \|\text{vec}(A)\|_1$ and $\|A\|_\infty = \|\text{vec}(A)\|_\infty$. Further, we define the entropy of a matrix $X \in \mathbb{R}_+^{n \times n}$ by

$$H(X) := -\sum_{i,j=1}^n X^{ij} \ln X^{ij}. \quad (4)$$

For two matrices $A, B$, we denote their Frobenius inner product by $\langle A, B \rangle$. We denote $\Delta^n := \{a \in \mathbb{R}_+^n : a^T \mathbf{1} = 1\}$ the probability simplex in $\mathbb{R}^n$. For $p, q \in \Delta^n$, we define the Kullback–Leibler divergence between $p$ and $q$ to be

$$KL(p\|q) := \sum_{i=1}^n p^i \ln \frac{p^i}{q^i}.$$

----

[4]A more general adaptive primal-dual method, which can solve problems both with linear equality and *inequality* constraints can be found in (Dvurechensky et al., 2017).

---

**Algorithm 1** Sinkhorn's Algorithm

**Input:** Accuracy $\varepsilon'$.
1: Set $k = 0$, $u_0 = v_0 = 0$
2: **repeat**
3:   **if** $k \mod 2 = 0$ **then**
4:     $u_{k+1} = u_k + \ln r - \ln(B(u_k, v_k)\mathbf{1})$
5:     $v_{k+1} = v_k$
6:   **else**
7:     $v_{k+1} = v_k + \ln c - \ln(B(u_k, v_k)^T\mathbf{1})$
8:     $u_{k+1} = u_k$
9:   **end if**
10:   $k = k + 1$
11: **until** $\|B(u_k, v_k)\mathbf{1} - r\|_1 + \|B(u_k, v_k)^T\mathbf{1} - c\|_1 \leq \varepsilon'$
**Output:** $B(u_k, v_k)$.

---

## 2. Sinkhorn's Algorithm

In this section, our goal is to refine the complexity analysis of the Sinkhorn's algorithm, then, based on this analysis, improve the existing complexity bound $O\left(\frac{n^2\|C\|_\infty^3 \log n}{\varepsilon^3}\right)$ for approximating the OT distance in the sense of (3) and obtain new complexity $O\left(\frac{n^2\|C\|_\infty^2 \log n}{\varepsilon^2}\right)$.

### 2.1. Improved Complexity of the Sinkhorn's Algorithm

We consider the Sinkhorn–Knopp algorithm listed as Algorithm 1, which solves (Cuturi, 2013), Lemma 2, the minimization problem

$$\min_{u,v \in \mathbb{R}^n}\left\{\psi(u, v) := \mathbf{1}^T B(u, v)\mathbf{1} - \langle u, r \rangle - \langle v, c \rangle\right\}, \quad (5)$$

where $K := e^{-C/\gamma}$ and $B(u, v) := \text{diag}\left(e^u\right) K \text{diag}\left(e^v\right)$, $\text{diag}(a)$ being the diagonal matrix with the vector $a$ on the diagonal. Problem (5) is equivalent to the dual to (2) with a particular choice $\mathcal{R}(X) = -H(X)$, see the derivation.

To improve the complexity of the Sinkhorn's algorithm, first, we obtain some bounds for the iterates $u_k$, $v_k$ and an optimal solution $(u^*, v^*)$ for (5). Then, using these bounds, for each iteration of the algorithm, we upper bound the objective value $\psi(u_k, v_k)$ by $\|B(u_k, v_k)\mathbf{1} - r\|_1 + \|B(u_k, v_k)^T\mathbf{1} - c\|_1$. Finally, the latter bound is used, to prove the main theorem of this subsection with new complexity result for the Sinkhorn's algorithm.

The following lemma provides the bounds for $u_k$, $v_k$, $u^*$ and $v^*$.

**Lemma 1.** *Let $k \geq 0$ and $u_k$, $v_k$ be generated by Algorithm 1 and $(u^*, v^*)$ be a solution of (5). Then*

$$\max_i u_k^i - \min_i u_k^i \leq R, \quad \max_j v_k^j - \min_j v_k^j \leq R, \quad (6)$$

$$\max_i (u^*)^i - \min_i (u^*)^i \leq R, \quad \max_j (v^*)^j - \min_j (v^*)^j \leq R,$$

*where*

$$R := -\ln\left(\nu \min_{i,j}\{r^i, c^j\}\right), \tag{7}$$

$$\nu := \min_{i,j} K^{ij} = e^{-\|C\|_\infty/\gamma}. \tag{8}$$

*Proof.* First, we prove the bound for $u_k$. Obviously, the stated inequality holds for $k = 0$. Let $k - 1$ be even. Then the variable $u$ is updated on the iteration $k - 1$ and $B(u_k, v_k)\mathbf{1} = r$ by the algorithm construction. Hence, for each $i \in [1, n]$, we have

$$e^{u_k^i}\nu\langle\mathbf{1}, e^{v_k}\rangle \leq \sum_j e^{u_k^i} K^{ij} e^{v_k^j} = [B(u_k, v_k)\mathbf{1}]_i = r_i \leq 1$$

$$\text{and} \quad \max_i u_k^i \leq -\ln\left(\nu\langle\mathbf{1}, e^{v_k}\rangle\right). \tag{9}$$

On the other hand, since $K^{ij} \leq 1$, for each $i \in [1, n]$,

$$e^{u_k^i}\langle\mathbf{1}, e^{v_k}\rangle \geq \sum_j e^{u_k^i} K^{ij} e^{v_k^j} = [B(u_k, v_k)\mathbf{1}]_i = r_i$$

$$\text{and} \quad \min_i u_k^i \geq \min_i \ln\left(\frac{r_i}{\langle\mathbf{1}, e^{v_k}\rangle}\right) = \ln\left(\frac{\min_i r^i}{\langle\mathbf{1}, e^{v_k}\rangle}\right).$$

The latter inequality and (9) give

$$\max_i u_k^i - \min_i u_k^i \leq -\ln\left(\nu \min_i r^i\right) \leq R.$$

Since the next iteration $k$, which is odd, updates the variable $v$ and leaves the variable $u$ unchanged, the obtained bound for $u_k$ holds for any $k \geq 0$. The bound in (6) for $v_k$ is proved in the same way. Finally, since $(u^*, v^*)$ is an optimal solution of (5), the gradient of the objective in (5) vanishes at this point. Hence, $B(u^*, v^*)\mathbf{1} = r$ and $B(u^*, v^*)^T\mathbf{1} = c$. Using these equalities and repeating the same arguments as in the proof of bounds for $u_k$ and $v_k$, we prove the bounds for $u^*$ and $v^*$.  □

The following lemma, for each iteration of Algorithm 1, relates the objective $\psi(u_k, v_k)$ in (5) and $\|B(u_k, v_k)\mathbf{1} - r\|_1 + \|B(u_k, v_k)^T\mathbf{1} - c\|_1$. To simplify derivations, we define

$$\widetilde{\psi}(u, v) := \psi(u, v) - \psi(u^*, v^*)$$
$$= \langle\mathbf{1}, B(u, v)\mathbf{1}\rangle - \langle\mathbf{1}, B(u^*, v^*)\mathbf{1}\rangle + \langle u^* - u, r\rangle + \langle v^* - v, c\rangle.$$

**Lemma 2.** *Let $k \geq 1$ and $u_k$, $v_k$ be generated by Algorithm 1. Then, denoting $B_k := B(u_k, v_k)$, we have*

$$\widetilde{\psi}(u_k, v_k) \leq R\left(\|B_k\mathbf{1} - r\|_1 + \|B_k^T\mathbf{1} - c\|_1\right).$$

*Proof.* Let us fix $k \geq 1$ and consider the convex function of $(\hat{u}, \hat{v})$

$$\langle\mathbf{1}, B(\hat{u}, \hat{v})\mathbf{1}\rangle - \langle\hat{u}, B(u_k, v_k)\mathbf{1}\rangle - \langle\hat{v}, B(u_k, v_k)^T\mathbf{1}\rangle.$$

Since its gradient vanishes at $(\hat{u}, \hat{v}) = (u_k, v_k)$, the point $(u_k, v_k)$ is its minimizer. Hence,

$$\widetilde{\psi}(u_k, v_k) = \left[\langle\mathbf{1}, B_k\mathbf{1}\rangle - \langle u_k, B_k\mathbf{1}\rangle - \langle v_k, B_k^T\mathbf{1}\rangle\right]$$
$$- \left[\langle\mathbf{1}, B(u^*, v^*)\mathbf{1}\rangle - \langle u^*, B_k\mathbf{1}\rangle - \langle v^*, B_k^T\mathbf{1}\rangle\right]$$
$$+ \langle u_k - u^*, B_k\mathbf{1} - r\rangle + \langle v_k - v^*, B_k^T\mathbf{1} - c\rangle$$
$$\leq \langle u_k - u^*, B_k\mathbf{1} - r\rangle + \langle v_k - v^*, B_k^T\mathbf{1} - c\rangle. \tag{10}$$

Next, we bound the r.h.s. of this inequality. Since, on each iteration of the Sinkhorn's algorithm, either $B_k\mathbf{1} = r$, or $B_k^T\mathbf{1} = c$, we have that $\langle\mathbf{1}, B_k\mathbf{1}\rangle = 1$ and $\langle\mathbf{1}, B_k\mathbf{1} - r\rangle = 0$. Taking $a = \frac{\max_i u_k^i + \min_i u_k^i}{2}$, by Hölder's inequality and Lemma 1, we obtain

$$\langle u_k, B_k\mathbf{1} - r\rangle = \langle u_k - a\mathbf{1}, B_k\mathbf{1} - r\rangle$$
$$\leq \|u_k - a\mathbf{1}\|_\infty \|B_k\mathbf{1} - r\|_1$$
$$= \frac{\max_i u_k^i - \min_i u_k^i}{2} \|B_k\mathbf{1} - r\|_1 \leq \frac{R}{2} \|B_k\mathbf{1} - r\|_1.$$

Using the same arguments, we bound $\langle -u^*, B_k\mathbf{1} - r\rangle$, $\langle v_k, B_k^T\mathbf{1} - c\rangle$ and $\langle -v^*, B_k^T\mathbf{1} - c\rangle$ in (10) and finish the proof of the lemma.  □

Now we are ready to improve the *iteration* complexity bound for the Sinkhorn's algorithm.

**Theorem 1.** *Algorithm 1 outputs a matrix $B(u_k, v_k)$ satisfying $\|B(u_k, v_k)\mathbf{1} - r\|_1 + \|B(u_k, v_k)^T\mathbf{1} - c\|_1 \leq \varepsilon'$ in the number of iterations $k$ satisfying*

$$k \leq 2 + \frac{4R}{\varepsilon'}.$$

*Proof.* Assume that $k \geq 1$ is even. As before, we denote $B_k = B(u_k, v_k)$. Since $\langle\mathbf{1}, B_k\mathbf{1}\rangle = \langle\mathbf{1}, B_{k+1}\mathbf{1}\rangle = 1$ and $v_{k+1} = v_k$, we have

$$\widetilde{\psi}(u_k, v_k) - \widetilde{\psi}(u_{k+1}, v_{k+1})$$
$$= \langle\mathbf{1}, B_k\mathbf{1}\rangle - \langle\mathbf{1}, B_{k+1}\mathbf{1}\rangle + \langle u_{k+1} - u_k, r\rangle + \langle v_{k+1} - v_k, c\rangle$$
$$= \langle r, u_{k+1} - u_k\rangle = \langle r, \ln r - \ln(B_k\mathbf{1})\rangle = KL(r\|B_k\mathbf{1})$$

By Pinsker's inequality and Lemma 2, since $B_k^T\mathbf{1} = c$, we obtain

$$\widetilde{\psi}(u_k, v_k) - \widetilde{\psi}(u_{k+1}, v_{k+1}) = KL\left(r\|B_k\mathbf{1}\right)$$
$$\geq \frac{1}{2}\|B_k\mathbf{1} - r\|_1^2 \geq \max\left\{\frac{\widetilde{\psi}(u_k, v_k)^2}{2R^2}, \frac{(\varepsilon')^2}{2}\right\}, \tag{11}$$

where we also used that, as soon as the stopping criterion is not yet fulfilled and $B_k^T\mathbf{1} = c$, $\|B_k\mathbf{1} - r\|_1^2 \geq (\varepsilon')^2$. The same inequality can be proved for the case of odd $k$.

Therefore (Nesterov, 2004), §2.1.5, for any $k \geq 1$,

$$\frac{\widetilde{\psi}(u_{k+1}, v_{k+1})}{2R^2} \leq \frac{\widetilde{\psi}(u_k, v_k)}{2R^2} - \left(\frac{\widetilde{\psi}(u_k, v_k)}{2R^2}\right)^2 \leq \frac{1}{k+\ell}, \quad (12)$$

where $\ell = \frac{2R^2}{\widetilde{\psi}(u_1, v_1)}$. Thus $k \leq 1 + \frac{2R^2}{\widetilde{\psi}(u_k, v_k)} - \frac{2R^2}{\widetilde{\psi}(u_1, v_1)}$. On the other hand,

$$\widetilde{\psi}(u_{k+m}, v_{k+m}) \leq \widetilde{\psi}(u_k, v_k) - \frac{(\varepsilon')^2 m}{2}, \quad k, m \geq 0. \quad (13)$$

To combine the two estimates (12) and (13), we consider a switching strategy, parametrized by number $s \in (0, \widetilde{\psi}(u_1, v_1)]$. First, using (12), we estimate the number of iterations to reduce $\widetilde{\psi}(u, v)$ from $\widetilde{\psi}(u_1, v_1)$ to $s$. Then, using (13), we estimate the number of iterations to reduce $\widetilde{\psi}(u, v)$ from $s$ to zero, keeping in mind that $\widetilde{\psi}(u, v) \geq 0$ by its definition. Minimizing the sum of these two estimates in $s \in (0, \widetilde{\psi}(u_1, v_1)]$, we conclude that the total number of iterations $k$ satisfies

$$k \leq \min_{0 < s \leq \widetilde{\psi}(u_1, v_1)} \left(2 + \frac{2R^2}{s} - \frac{2R^2}{\widetilde{\psi}(u_1, v_1)} + \frac{2s}{(\varepsilon')^2}\right)$$
$$= \begin{cases} 2 + \frac{4R}{\varepsilon'} - \frac{2R^2}{\widetilde{\psi}(u_1, v_1)}, & \widetilde{\psi}(u_1, v_1) \geq R\varepsilon', \\ 2 + \frac{2\widetilde{\psi}(u_1, v_1)}{(\varepsilon')^2}, & \widetilde{\psi}(u_1, v_1) < R\varepsilon'. \end{cases}$$

In both cases, we have $k \leq 2 + \frac{4R}{\varepsilon'}$. $\square$

The main innovation of our proof is the first component in the $\max$ in r.h.s. of (11), which follows from Lemma 2. On the contrary, (Altschuler et al., 2017) (see also (Chakrabarty & Khanna, 2018)) prove only the bound with the second component and, thus, obtain worse estimate for the number of Sinkhorn's iterations.

### 2.2. Complexity of OT Distance by Sinkhorn

Now we apply the result of the previous subsection to derive a complexity estimate for finding $\widehat{X} \in \mathcal{U}(r, c)$ satisfying (3). The procedure for approximating the OT distance by the Sinkhorn's algorithm is listed as Algorithm 2.

**Theorem 2.** *Algorithm 2 outputs $\widehat{X} \in \mathcal{U}(r, c)$ satisfying (3) in*

$$O\left(\frac{n^2 \|C\|_\infty^2 \ln n}{\varepsilon^2}\right) \quad arithmetic \ operations.$$

Before we prove the theorem, we compare our result with the best known in the literature, which is given by (Altschuler et al., 2017), Theorem 1: $O\left(\frac{n^2 \|C\|_\infty^3 \ln n}{\varepsilon^3}\right)$. As we see, our result has better dependence on $\varepsilon$ and $\|C\|_\infty$.

---

**Algorithm 2** Approximate OT by Sinkhorn

**Input:** Accuracy $\varepsilon$.
1: Set $\gamma = \frac{\varepsilon}{4 \ln n}$, $\varepsilon' = \frac{\varepsilon}{8\|C\|_\infty}$.
2: Find $\tilde{r}, \tilde{c} \in \Delta^n$ s.t. $\|\tilde{r} - r\|_1 \leq \varepsilon'/4$, $\|\tilde{c} - c\|_1 \leq \varepsilon'/4$ and $\min_i \tilde{r}^i \geq \varepsilon'/(8n)$, $\min_j \tilde{c}^j \geq \varepsilon'/(8n)$.
   E.g., $(\tilde{r}, \tilde{c}) = \left(1 - \frac{\varepsilon'}{8}\right)\left((r, c) + \frac{\varepsilon'}{n(8-\varepsilon')}(\mathbf{1}, \mathbf{1})\right)$.
3: Calculate $B$ by Algorithm 1 with marginals $\tilde{r}, \tilde{c}$ and accuracy $\varepsilon'/2$.
4: Find $\widehat{X}$ as the projection of $B$ on $\mathcal{U}(r, c)$ by Algorithm 2 in (Altschuler et al., 2017).

**Output:** $\widehat{X}$.

---

*Proof of Theorem 2.* Following the same steps as in the proof of Theorem 1 in (Altschuler et al., 2017), we obtain

$$\langle C, \widehat{X} \rangle \leq \langle C, X^* \rangle + 2\gamma \ln n + 4(\|B\mathbf{1} - r\|_1 + \|B^T\mathbf{1} - c\|_1)\|C\|_\infty, \quad (14)$$

where $\widehat{X}$ is the output of Algorithm 2, $X^*$ is a solution to the OT problem (3), and $B$ is the matrix obtained in step 3 of this Algorithm 2. At the same time, we have

$$\|B\mathbf{1} - r\|_1 + \|B^T\mathbf{1} - c\|_1$$
$$\leq \|B\mathbf{1} - \tilde{r}\|_1 + \|\tilde{r} - r\|_1 + \|B^T\mathbf{1} - \tilde{c}\|_1 + \|\tilde{c} - c\|_1 \leq \varepsilon'$$

Setting $\gamma = \frac{\varepsilon}{4 \ln n}$ and $\varepsilon' = \frac{\varepsilon}{8\|C\|_\infty}$, we obtain from the above inequality and (14) that $\widehat{X}$ satisfies inequality (3).

It remains to estimate the complexity of Algorithm 2. By Theorem 1, when $\varepsilon'$ is sufficiently small, the number of iterations of the Sinkhorn's algorithm in step 3 of Algorithm 2 is $O\left(\frac{R}{\varepsilon'}\right)$, where, according to (7) and (8),

$$R = -\ln\left(\nu \min_{i,j}\{\tilde{r}^i, \tilde{c}^j\}\right)$$
$$= -\ln\left(e^{-\|C\|_\infty/\gamma} \min_{i,j}\{\tilde{r}^i, \tilde{c}^j\}\right) \leq \frac{\|C\|_\infty}{\gamma} - \ln\left(\frac{\varepsilon'}{8n}\right).$$

Since $\gamma = \frac{\varepsilon}{4 \ln n}$ and $\varepsilon' = \frac{\varepsilon}{8\|C\|_\infty}$, we obtain that $R = O\left(\frac{\|C\|_\infty \ln n}{\varepsilon}\right)$. Inserting this into the estimate $k = O\left(\frac{R}{\varepsilon'}\right)$, we obtain that the total number of Sinkhorn's algorithm iterations is bounded by $O\left(\frac{\|C\|_\infty^2 \ln n}{\varepsilon^2}\right)$. Obviously, $\tilde{r}$ and $\tilde{c}$ in step 2 of Algorithm 2 can be found in $O(n)$ time. Since each iteration of the Sinkhorn's algorithm requires $O(n^2)$ arithmetic operations, the total complexity of Algorithm 2 is $O\left(\frac{n^2\|C\|_\infty^2 \ln n}{\varepsilon^2}\right)$. $\square$

Note that, as a byproduct, we obtained a theoretical justification of a commonly used in practice heuristic trick of changing zero values of measures $r, c$ to some small positive values.

## 3. Accelerated Gradient Descent

In this section, our goal is to propose a flexible algorithm for solving the regularized OT problem (2) with a general strongly convex regularizer and, based on this algorithm, obtain a complexity bound $\widetilde{O}\left(\min\left\{\frac{n^{9/4}}{\varepsilon}, \frac{n^2}{\varepsilon^2}\right\}\right)$ for approximating the OT distance in the sense of (3). To achieve this goal, we consider a general optimization problem, of which (2) is a particular case, and provide an Adaptive Primal-Dual Accelerated Gradient Descent (APDAGD) method for this problem together with its convergence rate. Finally, we apply this algorithm to the entropy-regularized OT problem and obtain the desired complexity.

### 3.1. General Problem and Algorithm

In this subsection, we consider the optimization problem

$$\min_{x \in Q \subseteq E} \{f(x) : Ax = b\}, \quad (15)$$

where $E$ is a finite-dimensional real vector space, $Q$ is a simple closed convex set, $A$ is a given linear operator from $E$ to some finite-dimensional real vector space $H$, $b \in H$ is given, $f(x)$ is a $\gamma$-strongly convex function on $Q$ with respect to some chosen norm $\|\cdot\|_E$ on $E$.

The Lagrange dual problem for (15), written as a minimization problem, is

$$\min_{\lambda \in H^*} \left\{\varphi(\lambda) := \langle \lambda, b \rangle + \max_{x \in Q}\left(-f(x) - \langle A^T\lambda, x \rangle\right)\right\}. \quad (16)$$

Note that $\nabla\varphi(\lambda) = b - Ax(\lambda)$ is Lipschitz-continuous (Nesterov, 2005)

$$\|\nabla\varphi(\lambda_1) - \nabla\varphi(\lambda_1)\|_H \le L\|\lambda_1 - \lambda_2\|_{H,*},$$

where $x(\lambda) = \arg\min_{x \in Q}\left(-f(x) - \langle A^T\lambda, x \rangle\right)$ and $L \le \frac{\|A\|^2_{E \to H}}{\gamma}$. This estimate can be pessimistic and our algorithm does not use it and adapts automatically to the local value of the Lipschitz constant.

We assume that the dual problem (16) has a solution and there exists some $R > 0$ such that $\|\lambda^*\|_2 \le R < +\infty$, where $\lambda^*$ is the solution to (16) with minimum value of $\|\lambda^*\|_2$. Note that the algorithm does not need any estimate of $R$ and the value $R$ is used only in the convergence analysis.

This algorithm can be considered as a primal-dual extension of accelerated mirror descent (Tseng, 2008; Lan et al., 2011). The difference to the literature consists in incorporating line-search and an online stopping criterion based only on the duality gap and constraints infeasibility. We provide a more detailed discussion in the supplementary material. Gradient methods for non-convex problems with line-search can be found in (Bogolubsky et al., 2016; Dvurechensky, 2017).

---

**Algorithm 3** Adaptive Primal-Dual Accelerated Gradient Descent (APDAGD)

**Input:** Accuracy $\varepsilon_f, \varepsilon_{eq} > 0$, initial estimate $L_0$ s.t. $0 < L_0 < 2L$.
1: Set $i_0 = k = 0$, $M_{-1} = L_0$, $\beta_0 = \alpha_0 = 0$, $\eta_0 = \zeta_0 = \lambda_0 = 0$.
2: **repeat** {Main iterate}
3:   **repeat** {Line search}
4:     Set $M_k = 2^{i_k-1}M_k$, find $\alpha_{k+1}$ s.t. $\beta_{k+1} := \beta_k + \alpha_{k+1} = M_k\alpha_{k+1}^2$. Set $\tau_k = \alpha_{k+1}/\beta_{k+1}$.
5:     $\lambda_{k+1} = \tau_k\zeta_k + (1 - \tau_k)\eta_k$.
6:     $\zeta_{k+1} = \zeta_k - \alpha_{k+1}\nabla\varphi(\lambda_{k+1})$.
7:     $\eta_{k+1} = \tau_k\zeta_{k+1} + (1 - \tau_k)\eta_k$.
8:   **until**

$$\varphi(\eta_{k+1}) \le \varphi(\lambda_{k+1}) + \langle\nabla\varphi(\lambda_{k+1}), \eta_{k+1} - \lambda_{k+1}\rangle$$
$$+ \frac{M_k}{2}\|\eta_{k+1} - \lambda_{k+1}\|_2^2.$$

9:   $\hat{x}_{k+1} = \tau_k x(\lambda_{k+1}) + (1 - \tau_k)\hat{x}_k$.
10:   Set $i_{k+1} = 0$, $k = k + 1$.
11: **until** $f(\hat{x}_{k+1}) + \varphi(\eta_{k+1}) \le \varepsilon_f$, $\|A\hat{x}_{k+1} - b\|_2 \le \varepsilon_{eq}$.
**Output:** $\hat{x}_{k+1}, \eta_{k+1}$.

---

**Theorem 3.** *Assume that the objective in the primal problem (15) is $\gamma$-strongly convex and that the dual solution $\lambda^*$ satisfies $\|\lambda^*\|_2 \le R$. Then, for $k \ge 1$, the points $\hat{x}_k, \eta_k$ in Algorithm 3 satisfy*

$$f(\hat{x}_k) - f^* \le f(\hat{x}_k) + \varphi(\eta_k) \le \frac{16\|A\|_{E \to H}^2 R^2}{\gamma k^2}, \quad (17)$$

$$\|A\hat{x}_k - b\|_2 \le \frac{16\|A\|_{E \to H}^2 R}{\gamma k^2}, \quad (18)$$

$$\|\hat{x}_k - x^*\|_E \le \frac{8}{k}\frac{\|A\|_{E \to H}R}{\gamma}, \quad (19)$$

*where $x^*$ and $f^*$ are respectively an optimal solution and the optimal value in (15). Moreover, the stopping criterion in step 11 is correctly defined.*

A stronger statement of the theorem and its proof can be found in the supplementary material.

Note that APDAGD is indeed flexible. For the case of entropy regularization, we set $f(X) = \langle C, X \rangle - \gamma H(X)$ and immediately get an algorithm to solve (2) since $-H(X)$ is strongly convex w.r.t. $\|\cdot\|_1$. For the case of Euclidean norm regularization, we set $f(X) = \langle C, X \rangle + \gamma\|X\|_2^2$ and obtain strong convexity w.r.t. the Euclidean norm. Other strongly convex regularizes are also suitable.

### 3.2. Complexity of OT Distance by APDAGD

Now we apply the result of the previous subsection to derive a complexity estimate for finding $\widehat{X} \in \mathcal{U}(r, c)$ satisfying

---

**Algorithm 4** Approximate OT by APDAGD

**Input:** Accuracy $\varepsilon$.
1: Set $\gamma = \frac{\varepsilon}{3 \ln n}$.
2: **for** $k = 1, 2, \dots$ **do**
3:     Make step of APDAGD and calculate $\widehat{X}_k$ and $\eta_k$.
4:     Find $\widehat{X}$ as the projection of $\widehat{X}_k$ on $\mathcal{U}(r, c)$ by Algorithm 2 in (Altschuler et al., 2017).
5:     **if** $\langle C, \widehat{X} - \widehat{X}_k \rangle \leq \frac{\varepsilon}{6}$ and $f(\hat{x}_k) + \varphi(\eta_k) \leq \frac{\varepsilon}{6}$ **then**
6:         Return $\widehat{X}$.
7:     **else**
8:         $k = k + 1$ and continue.
9:     **end if**
10: **end for**

---

(3). We use entropic regularization of problem (1) and consider the regularized problem (2) with the regularizer $\mathcal{R}(X) = -H(X)$, where $H(X)$ is given in (4). We define $E = \mathbb{R}^{n^2}$, $\| \cdot \|_E = \| \cdot \|_1$, and variable $x = \mathrm{vec}(X) \in \mathbb{R}^{n^2}$ to be the vector obtained from a matrix $X$ by writing each column of $X$ below the previous column. Also we set $f(x) = \langle C, X \rangle - \gamma H(X)$, $Q = \mathbb{R}^{n^2}_+$, $b^T = (r^T, c^T)$ and $A : \mathbb{R}^{n^2} \to \mathbb{R}^{2n}$ defined by the identity $(A \mathrm{vec}(X))^T = ((X\mathbf{1})^T, (X^T\mathbf{1})^T)$. With this setting, we solve problem (15) by our APDAGD. Let $\widehat{X}_k$ be defined by identity $\mathrm{vec}(\widehat{X}_k) = \hat{x}_k$, where $\hat{x}_k$ is generated by APDAGD. We also define $\widehat{X} \in \mathcal{U}(r, c)$ to be the projection of $\widehat{X}_k$ onto $\mathcal{U}(r, c)$ constructed by Algorithm 2 in (Altschuler et al., 2017). The pseudocode of our procedure for approximating the OT distance is listed as Algorithm 4.

**Theorem 4.** *Algorithm 4 outputs $\widehat{X} \in \mathcal{U}(r, c)$ satisfying (3) in*

$$O\left( \min\left\{ \frac{n^{9/4} \sqrt{R\|C\|_\infty \ln n}}{\varepsilon}, \frac{n^2 R\|C\|_\infty \ln n}{\varepsilon^2} \right\} \right) \tag{20}$$

*arithmetic operations.*

Before we prove the theorem, we compare our result with the best known in the literature, which is given by (Altschuler et al., 2017), Theorem 1: $O\left( \frac{n^2 \|C\|_\infty^3 \ln n}{\varepsilon^3} \right)$. As we see, our result in (20) has much better dependence on $\varepsilon$ and $\|C\|_\infty$, which comes for a reasonable price of $n^{1/4}$. We also underline that the complexity (20) obtained with the accelerated gradient descent Algorithm 4 has better dependence on $\varepsilon$ and $\|C\|_\infty$ than our improved bound for the Sinkhorn's algorithm given in Theorem 2: $O\left( \frac{n^2 \|C\|_\infty^2 \ln n}{\varepsilon^2} \right)$.

Importantly, similarly to the Sinkhorn's algorithm, our APDAGD algorithm can be parallelized, and efficiently implemented when the Sinkhorn kernel matrix $\exp(-C/\gamma)$ is easy to apply (Solomon et al., 2015), e.g. the measures are supported on regular grids and $C$ is given by squared

Euclidean distance. See the supplementary material for the details.

*Proof of Theorem 4.* Let $X^*$ be the solution of the OT problem (1) and $X^*_\gamma$ be the solution of the regularized problem (2). Then, we have

$$\langle C, \widehat{X} \rangle = \langle C, X^* \rangle + \langle C, X^*_\gamma - X^* \rangle \\ + \langle C, \widehat{X}_k - X^*_\gamma \rangle + \langle C, \widehat{X} - \widehat{X}_k \rangle. \tag{21}$$

Now we estimate the second and third term in the r.h.s. Since, for any $X \in \mathcal{U}(r, c)$, $-H(X) \in [-2 \ln n, 0]$, we have

$$\langle C, X^*_\gamma - X^* \rangle = \min_{X \in \mathcal{U}(r,c)} \{ \langle C, X \rangle - \gamma H(X) \} \\ - \min_{X \in \mathcal{U}(r,c)} \langle C, X \rangle \leq 0. \tag{22}$$

Further, since APDAGD solves problem (15) with $f(x) = \langle C, X \rangle - \gamma H(X)$ and $X^*_\gamma$ is the solution, we have

$$\langle C, \widehat{X}_k - X^*_\gamma \rangle = (\langle C, \widehat{X}_k \rangle - \gamma H(\widehat{X}_k)) \\ - (\langle C, X^*_\gamma \rangle - \gamma H(X^*_\gamma)) + \gamma (H(\widehat{X}_k) - H(X^*_\gamma)) \\ \overset{(17)}{\leq} f(\hat{x}_k) + \varphi(\eta_k) + 2\gamma \ln n, \tag{23}$$

where we used again that $-H(X) \in [-2 \ln n, 0]$ for $X \in \mathcal{U}(r, c)$. Combining (21), (22) and (23), we obtain

$$\langle C, \widehat{X} \rangle \leq \langle C, X^* \rangle + \langle C, \widehat{X} - \widehat{X}_k \rangle \\ + f(\hat{x}_k) + \varphi(\eta_k) + 2\gamma \ln n. \tag{24}$$

We immediately see that, when the stopping criterion in step 5 of Algorithm 4 is fulfilled, the output $\widehat{X} \in \mathcal{U}(r, c)$ satisfies (3).

It remains to obtain the complexity bound. First, we estimate the number of iterations in Algorithm 4 to guarantee $\langle C, \widehat{X} - \widehat{X}_k \rangle \leq \frac{\varepsilon}{6}$ and, after that, estimate the number of iterations to guarantee $f(\hat{x}_k) + \varphi(\eta_k) \leq \frac{\varepsilon}{6}$. By Hölder's inequality, we have $\langle C, \widehat{X} - \widehat{X}_k \rangle \leq \|C\|_\infty \|\widehat{X} - \widehat{X}_k\|_1$. By Lemma 7 in (Altschuler et al., 2017),

$$\|\widehat{X} - \widehat{X}_k\|_1 \leq 2 \left( \|\widehat{X}_k \mathbf{1} - r\|_1 + \|\widehat{X}_k^T \mathbf{1} - c\|_1 \right). \tag{25}$$

Next, we obtain two estimates for the r.h.s of this inequality. First, by the definition of the operator $A$ and vector $b$,

$$\|\widehat{X}_k \mathbf{1} - r\|_1 + \|\widehat{X}_k^T \mathbf{1} - c\|_1 \leq \sqrt{2n} \|A\mathrm{vec}(\widehat{X}_k) - b\|_2 \\ \overset{(18)}{\leq} \frac{16R\|A\|_{E \to H}^2 \sqrt{2n}}{\gamma k^2} \leq \frac{32R\sqrt{2n}}{\gamma k^2}. \tag{26}$$

Here we used the choice of the norm $\| \cdot \|_1$ in $E = \mathbb{R}^{n^2}$ and the norm $\| \cdot \|_2$ in $H = \mathbb{R}^{2n}$. Indeed, in this setting $\|A\|_{E \to H} = \|A\|_{1 \to 2}$ and this norm is equal to the maximum Euclidean norm of a column of $A$. By definition, each column of $A$ contains only two non-zero elements, which are equal to one. Hence, $\|A\|_{1 \to 2} = \sqrt{2}$.

Second, since $X_\gamma^* \in \mathcal{U}(r, c)$, we have

$$\|\widehat{X}_k \mathbf{1} - r\|_1 = \|(\widehat{X}_k - X_\gamma^*)\mathbf{1}\|_1 \leq \|\widehat{X}_k - X_\gamma^*\|_1$$

and a similar estimate for $\|\widehat{X}_k^T \mathbf{1} - c\|_1$. Combining these estimates with (19) and an estimate for $\|A\|_{E \to H}$, we obtain

$$\|\widehat{X}_k \mathbf{1} - r\|_1 + \|\widehat{X}_k^T \mathbf{1} - c\|_1 \leq 2\|\widehat{X}_k - X_\gamma^*\|_1 \overset{(19)}{\leq} \frac{16 R \sqrt{2}}{\gamma k}. \tag{27}$$

Combining (25), (26) and (27), we obtain

$$\langle C, \widehat{X} - \widehat{X}_k \rangle \leq 2\|C\|_\infty \min \left\{ \frac{32 R \sqrt{2n}}{\gamma k^2}, \frac{16 R \sqrt{2}}{\gamma k} \right\}.$$

Setting $\gamma = \frac{\varepsilon}{3 \ln n}$, we have that, to obtain $\langle C, \widehat{X} - \widehat{X}_k \rangle \leq \frac{\varepsilon}{6}$, it is sufficient to choose

$$k = O \left( \min \left\{ \frac{n^{1/4} \sqrt{R\|C\|_\infty \ln n}}{\varepsilon}, \frac{R\|C\|_\infty \ln n}{\varepsilon^2} \right\} \right). \tag{28}$$

At the same time, since $\|A\|_{E \to H} = \sqrt{2}$,

$$f(\hat{x}_k) + \varphi(\eta_k) \overset{(17)}{\leq} \frac{32 R^2}{\gamma k^2}.$$

Since we set $\gamma = \frac{\varepsilon}{3 \ln n}$, we conclude that in order to obtain $f(\hat{x}_k) + \varphi(\eta_k) \leq \frac{\varepsilon}{6}$, it is sufficient to choose

$$k = O \left( \frac{R \sqrt{\ln n}}{\varepsilon} \right). \tag{29}$$

To estimate the total number of iterations, we should take maximum of (28) and (29). Normalizing the cost matrix $C$, we can set $\|C\|_\infty = 1$. At the same time, as one can see from (5), the change of the dual variables $u \to u + t\mathbf{1}$, $v \to v - t\mathbf{1}$, for any $t \in \mathbb{R}$ does not change the value of the dual objective. Thus, without loss of generality, we can set $R \leq 1$. Hence, the maximum of (28) and (29) is attained by (28).

Since each iteration of APDAGD uses only operations with matrices of the size $n \times n$ and vectors of the size $2n$, each iteration requires $O(n^2)$ arithmetic operations. At the same time, according to Lemma 7 in (Altschuler et al., 2017), the complexity of projecting $\widehat{X}_k$ on $\mathcal{U}(r, c)$ by their Algorithm 2 is $O(n^2)$. Thus, to obtain the total complexity of Algorithm 4 as in the Theorem statement, we just multiply (28) by $n^2$. □

## 4. Experiments

In this section, we provide an empirical illustration of the work of Algorithm 2 and Algorithm 4. We run experiments on randomly chosen real images from the MNIST dataset. By default, this dataset contains images of handwritten digits of the size 28 by 28 pixels. To understand the dependence on the number of pixels $n$, we resize MNIST images to be images of $28 \cdot s$ by $28 \cdot s$ pixels, where $s$ is an integer. We change all the zero elements in the measures, representing these images, to $10^{-6}$ and, then, normalize them, so that they sum up to one. As we show in the proof of Theorem 2, small perturbations of the vectors $r$ and $c$ (see step 2 of Alg. 2) do not influence much the theoretical guarantees for the Sinkhorn's algorithm approach. By similar arguments, these changes do not influence much the APDAGD approach. We run Algorithm 2 until the stopping criterion $\|B\mathbf{1} - r\|_1 + \|B^T \mathbf{1} - c\|_1 \leq \frac{\varepsilon}{8\|C\|_\infty}$ is fulfilled. As we can see from the proof of Theorem 4, the inequality $f(\hat{x}_k) + \varphi(\eta_k) \leq \frac{\varepsilon}{6}$ is fulfilled faster than $\langle C, \widehat{X} - \widehat{X}_k \rangle \leq \frac{\varepsilon}{6}$. Thus, we run Algorithm 4 until the latter inequality is fulfilled. To understand the dependence on $\varepsilon$, we choose several values of accuracy $\varepsilon \in [0.025, 0.12]$ and $s = 1$. For each value, we randomly choose 10 pairs of images, run Algorithm 2 and Algorithm 4, and average the results. It is worth noting that, in practice, the working time of Algorithm 2 is approximately proportional to $\frac{1}{\varepsilon}$. The reason could be in a pessimistic theoretical bound $R$ in Lemma 1. Figure 1 (left) illustrates the working time of two algorithms for different $\varepsilon$. To understand the dependence on $n$, we choose accuracy $\varepsilon = 0.1$ and several values of $s \in [1, 8]$. For each value, we randomly choose 5 pairs of images, run Algorithm 2 and Algorithm 4, and average the results. Figure 1 (right) illustrates the working time of two algorithms for different $n$.

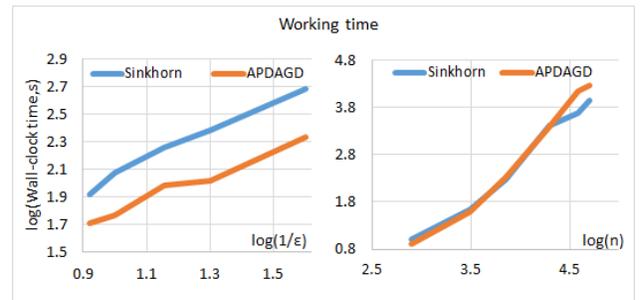

*Figure 1.* Comparison of working time of Algorithm 2 (Sinkhorn's algorithm) and Algorithm 4 (APDAGD).

## 5. Conclusion

We analyze two algorithms for approximating the general OT distances between two discrete distributions. Our first algorithm is based on the entropic regularization of the OT problem and Sinkhorn's algorithm. We prove the complexity bound $\widetilde{O}\left(\frac{n^2}{\varepsilon^2}\right)$ arithmetic operations. The second algorithm is based on the entropic regularization of the OT problem and our novel Adaptive Primal-Dual Accelerated Gradient method. We obtain the complexity $\widetilde{O}\left(\min\left\{\frac{n^{9/4}}{\varepsilon},\frac{n^2}{\varepsilon^2}\right\}\right)$ arithmetic operations for this algorithm. Both complexity bounds are better than the state-of-the-art result given by $\widetilde{O}\left(\frac{n^2}{\varepsilon^3}\right)$. Our APDAGD can be of a separate interest for solving strongly convex problems with linear constraints, since it is not specific to the entropic regularization, incorporates a line-search strategy and has an accelerated rate of convergence.

## Acknowledgements

We are grateful to the anonymous reviewers for their suggestions on how the paper can be improved, Benjamin Stemper for his help with proofreading the text, Sergey Omelchenko for the discussions of the experimental part of the paper, Alexander Tiurin for the discussions of the theoretical part of the paper. This work has been funded by the Russian Academic Excellence Project '5-100'. Results of Sections 3 and 4 have been obtained under support of the Russian Science Foundation (project 17-11-01027).

## References

Allen-Zhu, Z., Li, Y., Oliveira, R., and Wigderson, A. Much faster algorithms for matrix scaling. In *2017 IEEE 58th Annual Symposium on Foundations of Computer Science (FOCS)*, pp. 890–901, 2017. arXiv:1704.02315.

Altschuler, J., Weed, J., and Rigollet, P. Near-linear time approximation algorithms for optimal transport via sinkhorn iteration. In Guyon, I., Luxburg, U. V., Bengio, S., Wallach, H., Fergus, R., Vishwanathan, S., and Garnett, R. (eds.), *Advances in Neural Information Processing Systems 30*, pp. 1961–1971. Curran Associates, Inc., 2017. arXiv:1705.09634.

Anikin, A. S., Gasnikov, A. V., Dvurechensky, P. E., Tyurin, A. I., and Chernov, A. V. Dual approaches to the minimization of strongly convex functionals with a simple structure under affine constraints. *Computational Mathematics and Mathematical Physics*, 57(8):1262–1276, 2017.

Arjovsky, M., Chintala, S., and Bottou, L. Wasserstein GAN. *arXiv:1701.07875*, 2017.

Beck, A. and Teboulle, M. A fast dual proximal gradient algorithm for convex minimization and applications. *Operations Research Letters*, 42(1):1 – 6, 2014.

Bigot, J., Gouet, R., Klein, T., and López, A. Geodesic PCA in the wasserstein space by convex PCA. *Ann. Inst. H. Poincaré Probab. Statist.*, 53(1):1–26, 02 2017.

Blondel, M., Seguy, V., and Rolet, A. Smooth and sparse optimal transport. *arXiv:1710.06276*, 2017.

Bogolubsky, L., Dvurechensky, P., Gasnikov, A., Gusev, G., Nesterov, Y., Raigorodskii, A. M., Tikhonov, A., and Zhukovskii, M. Learning supervised pagerank with gradient-based and gradient-free optimization methods. In Lee, D. D., Sugiyama, M., Luxburg, U. V., Guyon, I., and Garnett, R. (eds.), *Advances in Neural Information Processing Systems 29*, pp. 4914–4922. Curran Associates, Inc., 2016. arXiv:1603.00717.

Bregman, L. Proof of the convergence of Sheleikhovskii's method for a problem with transportation constraints. *USSR Computational Mathematics and Mathematical Physics*, 7(1):191 – 204, 1967.

Chakrabarty, D. and Khanna, S. Better and simpler error analysis of the sinkhorn-knopp algorithm for matrix scaling. *arXiv:1801.02790*, 2018.

Chambolle, A. and Pock, T. A first-order primal-dual algorithm for convex problems with applications to imaging. *Journal of Mathematical Imaging and Vision*, 40(1):120–145, 2011.

Chernov, A., Dvurechensky, P., and Gasnikov, A. Fast primal-dual gradient method for strongly convex minimization problems with linear constraints. In Kochetov, Y., Khachay, M., Beresnev, V., Nurminski, E., and Pardalos, P. (eds.), *Discrete Optimization and Operations Research: 9th International Conference, DOOR 2016, Vladivostok, Russia, September 19-23, 2016, Proceedings*, pp. 391–403. Springer International Publishing, 2016.

Cohen, M. B., Madry, A., Tsipras, D., and Vladu, A. Matrix scaling and balancing via box constrained newton's method and interior point methods. In *2017 IEEE 58th Annual Symposium on Foundations of Computer Science (FOCS)*, pp. 902–913, 2017. arXiv:1704.02310.

Courty, N., Flamary, R., Tuia, D., and Rakotomamonjy, A. Optimal transport for domain adaptation. *IEEE Transactions on Pattern Analysis & Machine Intelligence*, 39(9):1853–1865, 2017.

Cuturi, M. Sinkhorn distances: Lightspeed computation of optimal transport. In Burges, C. J. C., Bottou, L., Welling, M., Ghahramani, Z., and Weinberger, K. Q. (eds.), *Advances in Neural Information Processing Systems 26*, pp. 2292–2300. Curran Associates, Inc., 2013.

Cuturi, M. and Peyré, G. A smoothed dual approach for variational wasserstein problems. *SIAM Journal on Imaging Sciences*, 9(1): 320–343, 2016.

Dvurechensky, P. Gradient method with inexact oracle for composite non-convex optimization. *arXiv:1703.09180*, 2017.

Dvurechensky, P., Gasnikov, A., Gasnikova, E., Matsievsky, S., Rodomanov, A., and Usik, I. Primal-dual method for searching equilibrium in hierarchical congestion population games. In *Supplementary Proceedings of the 9th International Conference on Discrete Optimization and Operations Research and Scientific School (DOOR 2016) Vladivostok, Russia, September 19 - 23, 2016*, pp. 584–595, 2016. arXiv:1606.08988.

Dvurechensky, P., Gasnikov, A., Omelchenko, S., and Tiurin, A. Adaptive similar triangles method: a stable alternative to sinkhorn's algorithm for regularized optimal transport. *arXiv:1706.07622*, 2017.

Ebert, J., Spokoiny, V., and Suvorikova, A. Construction of non-asymptotic confidence sets in 2-Wasserstein space. *arXiv:1703.03658*, 2017.

Essid, M. and Solomon, J. Quadratically-regularized optimal transport on graphs. *arXiv:1704.08200*, 2017.

Gasnikov, A. V., Gasnikova, E. V., Nesterov, Y. E., and Chernov, A. V. Efficient numerical methods for entropy-linear programming problems. *Computational Mathematics and Mathematical Physics*, 56(4):514–524, 2016.

Genevay, A., Cuturi, M., Peyré, G., and Bach, F. Stochastic optimization for large-scale optimal transport. In Lee, D. D., Sugiyama, M., Luxburg, U. V., Guyon, I., and Garnett, R. (eds.), *Advances in Neural Information Processing Systems 29*, pp. 3440–3448. Curran Associates, Inc., 2016.

Ho, N., Nguyen, X., Yurochkin, M., Bui, H. H., Huynh, V., and Phung, D. Multilevel clustering via Wasserstein means. In Precup, D. and Teh, Y. W. (eds.), *Proceedings of the 34th International Conference on Machine Learning*, volume 70, pp. 1501–1509. PMLR, 2017.

Kalantari, B. and Khachiyan, L. On the rate of convergence of deterministic and randomized ras matrix scaling algorithms. *Oper. Res. Lett.*, 14(5):237–244, 1993.

Kantorovich, L. On the translocation of masses. *Doklady Acad. Sci. USSR (N.S.)*, 37:199–201, 1942.

Kolouri, S., Park, S. R., Thorpe, M., Slepcev, D., and Rohde, G. K. Optimal mass transport: Signal processing and machine-learning applications. *IEEE Signal Processing Magazine*, 34 (4):43–59, 2017.

Kusner, M. J., Sun, Y., Kolkin, N. I., and Weinberger, K. Q. From word embeddings to document distances. In *Proceedings of the 32nd International Conference on International Conference on Machine Learning - Volume 37*, ICML'15, pp. 957–966. PMLR, 2015.

Lan, G., Lu, Z., and Monteiro, R. D. C. Primal-dual first-order methods with $O(1/\varepsilon)$ iteration-complexity for cone programming. *Mathematical Programming*, 126(1):1–29, 2011.

Lee, Y. T. and Sidford, A. Path finding methods for linear programming: Solving linear programs in $\tilde{O}(\sqrt{\text{rank}})$ iterations and faster algorithms for maximum flow. In *2014 IEEE 55th Annual Symposium on Foundations of Computer Science*, pp. 424–433, 2014.

Li, J., Wu, Z., Wu, C., Long, Q., and Wang, X. An inexact dual fast gradient-projection method for separable convex optimization with linear coupled constraints. *Journal of Optimization Theory and Applications*, 168(1):153–171, 2016.

Malitsky, Y. and Pock, T. A first-order primal-dual algorithm with linesearch. *arXiv:1608.08883*, 2016.

Nesterov, Y. *Introductory Lectures on Convex Optimization: a basic course*. Kluwer Academic Publishers, Massachusetts, 2004.

Nesterov, Y. Smooth minimization of non-smooth functions. *Mathematical Programming*, 103(1):127–152, 2005.

Ouyang, Y., Chen, Y., Lan, G., and Eduardo Pasiliao, J. An accelerated linearized alternating direction method of multipliers. *SIAM Journal on Imaging Sciences*, 8(1):644–681, 2015.

Panaretos, V. M. and Zemel, Y. Amplitude and phase variation of point processes. *Ann. Statist.*, 44(2):771–812, 2016.

Patrascu, A., Necoara, I., and Findeisen, R. Rate of convergence analysis of a dual fast gradient method for general convex optimization. In *2015 54th IEEE Conference on Decision and Control (CDC)*, pp. 3311–3316, 2015.

Pele, O. and Werman, M. Fast and robust Earth Mover's Distances. In *2009 IEEE 12th International Conference on Computer Vision*, pp. 460–467, 2009.

Pitié, F., Kokaram, A. C., and Dahyot, R. Automated colour grading using colour distribution transfer. *Computer Vision and Image Understanding*, 107(1):123 – 137, 2007.

Rubner, Y., Tomasi, C., and Guibas, L. J. The earth mover's distance as a metric for image retrieval. *International journal of computer vision*, 40(2):99–121, 2000.

Sandler, R. and Lindenbaum, M. Nonnegative matrix factorization with earth mover's distance metric for image analysis. *IEEE Transactions on Pattern Analysis and Machine Intelligence*, 33 (8):1590–1602, 2011.

Schmitzer, B. Stabilized sparse scaling algorithms for entropy regularized transport problems. *arXiv:1610.06519*, 2016.

Sinkhorn, R. Diagonal equivalence to matrices with prescribed row and column sums. II. *Proc. Amer. Math. Soc.*, 45:195–198, 1974.

Solomon, J., Rustamov, R. M., Guibas, L., and Butscher, A. Wasserstein propagation for semi-supervised learning. In *Proceedings of the 31st International Conference on International Conference on Machine Learning - Volume 32*, ICML'14, pp. I–306–I–314. PMLR, 2014.

Solomon, J., de Goes, F., Peyré, G., Cuturi, M., Butscher, A., Nguyen, A., Du, T., and Guibas, L. Convolutional wasserstein distances: Efficient optimal transportation on geometric domains. *ACM Trans. Graph.*, 34(4):66:1–66:11, 2015.

Tran-Dinh, Q. and Cevher, V. Constrained convex minimization via model-based excessive gap. In *Proceedings of the 27th International Conference on Neural Information Processing Systems*, NIPS'14, pp. 721–729, Cambridge, MA, USA, 2014. MIT Press.

Tseng, P. On accelerated proximal gradient methods for convex-concave optimization. Technical report, MIT, 2008. URL http://www.mit.edu/~dimitrib/PTseng/papers/apgm.pdf.

Villani, C. *Optimal transport: old and new*, volume 338. Springer Science & Business Media, 2008.

Werman, M., Peleg, S., and Rosenfeld, A. A distance metric for multidimensional histograms. *Computer Vision, Graphics, and Image Processing*, 32(3):328 – 336, 1985.

Xu, Y. Accelerated first-order primal-dual proximal methods for linearly constrained composite convex programming. *arXiv:1606.09155*, 2016.

Yurtsever, A., Tran-Dinh, Q., and Cevher, V. A universal primal-dual convex optimization framework. In *Proceedings of the 28th International Conference on Neural Information Processing Systems*, NIPS'15, pp. 3150–3158, Cambridge, MA, USA, 2015. MIT Press.

In this document, we provide some details on the dual problem for the regularized OT problem, which is used in the analysis of the Sinkhorn's algorithm, details on the efficient implementation of our APDAGD algorithm for the case of the Sinkhorn's kernel being easy to apply, e.g. when the measures are supported on regular grids and $C$ is given by squared Euclidean distance. Finally, we provide the missing proofs for the APDAGD-based approach. This is a separate document and, if not explicitly stated, all the references refer to formulas, algorithms, lemmas and theorems in this document.

# 1 Details for the Sinkhorn's Algorithm Approach

Below we provide the derivation of the dual problem for the regularized OT problem, which is used in Section 2.

$$
\min_{X \in \mathcal{U}(r,c)} \langle C, X \rangle + \gamma \langle X, \ln X \rangle = \min_{X \in \mathbb{R}_+^{n \times n}} \max_{y,z \in \mathbb{R}^n} \langle C, X \rangle + \gamma \langle X, \ln X \rangle + \langle y, X\mathbf{1} - r \rangle + \langle z, X^T\mathbf{1} - c \rangle
$$

$$
= \max_{y,z \in \mathbb{R}^n} -\langle y, r \rangle - \langle z, c \rangle + \min_{X^{ij} \geq 0} \sum_{i,j=1}^n X^{ij}(C^{ij} + \gamma \ln X^{ij} + y^i + z^j)
$$

$$
\left[ X^{ij} = \exp\left( -\frac{1}{\gamma}(y^i + z^j + C^{ij}) - 1 \right) \right]
$$

$$
\max_{y,z \in \mathbb{R}^n} -\langle y, r \rangle - \langle z, c \rangle - \gamma \sum_{i,j=1}^n \exp\left( -\frac{1}{\gamma}(y^i + z^j + C^{ij}) - 1 \right).
$$

(1)

Changing the variables $u = -y/\gamma - 1/2$, $v = -z/\gamma - 1/2$, disregarding the constant term $-1$ and dividing the objective by $-\gamma$, we obtain the dual problem considered in Section 2.

# 2 Details for the Accelerated Gradient Descent Approach

## 2.1 Efficient Implementation for Entropic Regularization

In this subsection, we show that the steps of APDAGD can be written in terms of the multiplication of the Sinkhorn kernel matrix $K = \exp(-C/\gamma)$ by a vector. Then, similarly to the Sinkhorn's algorithm the step of APDAGD can be performed faster, if this kernel is easy to apply, e.g. the measures are supported on regular grids and $C$ is given by squared Euclidean distance.

In the particular case of solving the entropy-regularized OT problem by APDAGD, we have $f(x) = \langle C, X \rangle + \gamma \langle X, \ln X \rangle$, $Q = \mathbb{R}_+^{n^2}$, $b^T = (r^T, c^T)$ and $A : \mathbb{R}^{n^2} \to \mathbb{R}^{2n}$ defined by the identity $(A \operatorname{vec}(X))^T = ((X\mathbf{1})^T, (X^T\mathbf{1})^T)$, $\lambda = (y, z)$, where $y$, $z$ are the dual variables, defined in (1). Then, $x(\lambda) = \arg\min_{x \in Q} \left( -f(x) - \langle A^T \lambda, x \rangle \right)$ satisfies $x(\lambda) = \operatorname{vec}(X(y, z))$, where, due to (1), $X(y, z) = e^{-1} \cdot \operatorname{diag}\left( e^{y/\gamma} \right) K \operatorname{diag}\left( e^{z/\gamma} \right)$. At the same time,

$$\nabla \varphi(\lambda) = b - Ax(\lambda) = \begin{pmatrix} r - X(y, z)\mathbf{1} \\ c - X(y, z)^T \mathbf{1} \end{pmatrix}$$

Hence, in order to calculate the gradient $\nabla \varphi(\lambda)$ in step 6 of APDAGD algorithm, one needs to apply the kernel $K$ to the vector $e^{y/\gamma}$ and to the vector $e^{z/\gamma}$, which is easy in the considered situation. Step 8 also involves the kernel $K$ for calculating the value $\varphi(\lambda)$. This is also easy since $\varphi(\lambda)$ can be written as

$$\varphi(\lambda) = \gamma \mathbf{1}^T X(y, z) \mathbf{1} + \langle y, r \rangle + \langle z, c \rangle = \frac{\gamma}{e} e^{y/\gamma} K e^{z/\gamma} + \langle y, r \rangle + \langle z, c \rangle.$$

Again $K$ is used only through matrix-vector multiplication. Other steps operate with dual variables $\lambda$, $\eta$, $\zeta \in \mathbb{R}^{2n}$ and, thus are also easy. Overall, APDAGD uses the same set of operations as the Sinkhorn's algorithm and, thus can also be implemented in parallel framework.

## 2.2 Adaptive Primal-Dual Accelerated Gradient Descent (AP-DAGD) Convergence Analysis

We provide the missing convergence rate proofs for the Adaptive Primal-Dual Accelerated Gradient Descent method for constrained convex optimization problem, which was considered in Section 3 of the main part of the paper.

We consider the problem

$$(P_1) \qquad \min_{x \in Q \subseteq E} \{f(x) : Ax = b\},$$

where $f(x)$ is a $\gamma$-strongly convex function on $Q$. The Lagrange dual problem to Problem $(P_1)$ in a form of a minimization problem is

$$\min_{\lambda \in \Lambda} \left\{ \varphi(\lambda) := \langle \lambda, b \rangle + \max_{x \in Q} \left( -f(x) - \langle A^T \lambda, x \rangle \right) \right\},$$

where $\Lambda$ is the space of Lagrange multipliers and, hence is *unbounded*.

Our Adaptive Primal-Dual Accelerated Gradient Descent method can be considered as an Adaptive Accelerated Gradient Descent applied to the dual problem and supplied with a procedure to reconstruct the primal iterate. Since it can be of independent interest, we first, in subsection 2.3 consider Adaptive Accelerated Gradient Descent method for a general convex optimization problem and prove in Theorem 1 its convergence rate in a primal-dual friendly fashion. Then, in subsection 2.4 we use this result to analyze our Adaptive Primal-Dual Accelerated Gradient Descent method.

## 2.3 Adaptive Accelerated Gradient Descent for Convex Optimization

In this section, we consider a general optimization problem

$$\min_{\lambda \in \Lambda} \varphi(\lambda), \tag{2}$$

where $\Lambda \in H^*$ is a closed convex, generally speaking, *unbounded*, set, $\varphi(\lambda)$ is a convex function with $L$-Lipschitz-continuous gradient, i.e.

$$\varphi(\eta) \leq \varphi(\lambda) + \langle \nabla\varphi(\lambda), \eta - \lambda \rangle + \frac{L}{2} \|\eta - \lambda\|_{H,*}^2, \quad \forall \eta, \lambda \in H^*. \tag{3}$$

### 2.3.1 Proximal Setup

In this subsection, we introduce *proximal setup*, which is usually used in proximal gradient methods, see e.g. Ben-Tal and Nemirovski [2015]. We choose some norm $\|\cdot\|$ on the space of vectors $\lambda$ and a *prox-function* $d(\lambda)$ which is continuous, convex on $\Lambda$ and

1. admits a continuous in $\lambda \in \Lambda^0$ selection of subgradients $\nabla d(\lambda)$, where $\lambda \in \Lambda^0 \subseteq \Lambda$ is the set of all $\lambda$, where $\nabla d(\lambda)$ exists;

2. is 1-strongly convex on $\Lambda$ with respect to $\|\cdot\|$, i.e., for any $\lambda \in \Lambda^0, \eta \in \Lambda$, $d(\eta) - d(\lambda) - \langle \nabla d(\lambda), \eta - \lambda \rangle \geq \frac{1}{2} \|\eta - \lambda\|^2$.

We define also the corresponding *Bregman divergence* $V[\zeta](\lambda) := d(\lambda) - d(\zeta) - \langle \nabla d(\zeta), \lambda - \zeta \rangle$, $\lambda \in \Lambda, \zeta \in \Lambda^0$. It is easy to see that

$$V[\zeta](\lambda) \geq \frac{1}{2} \|\lambda - \zeta\|^2, \quad \lambda \in \Lambda, \zeta \in \Lambda^0. \tag{4}$$

Standard proximal setups, i.e. Euclidean, entropy, $\ell_1/\ell_2$, simplex, nuclear norm, spectahedron can be found in Ben-Tal and Nemirovski [2015].

### 2.3.2 Algorithm and Complexity Analysis

In this subsection, we present Adaptive Accelerated Gradient Descent (see Algorithm 1 below) and prove its convergence rate theorem. Our algorithm in its form is very close to [Tseng, 2008, Alg.1] and [Lan et al., 2011, "Variant of Nesterov's algorithm"]. Nevertheless, the algorithms in those two papers assume the Lipschitz constant $L$ to be known and explicitly use it in the algorithm. Our algorithm is free of this drawback. Another distinction of our algorithm is that we prove convergence rate in a primal-dual-friendly manner. As we show in subsection 2.4, this allows us to apply our AAGD to the Lagrange dual problem for $(P_1)$, and reconstruct also primal iterates. In his paper, Tseng obtains primal-dual rates, but only for the case of bounded set $\Lambda$. In our case this analysis is inapplicable since the feasible set of the Lagrange dual problem is unbounded. Lan, Lu and Monteiro, consider a special problem of minimizing a linear function and do not prove primal-dual rates for their variant of Nesterov's algorithm.

We denote by $\eta_k, \zeta_k, \lambda_k$ three sequences of iterates of the algorithm and by $\alpha_k, \beta_k$ two sequences of numbers. The convergence rate is proved for the points $\eta_k$.

**Lemma 1.** *Algorithm 1 is defined correctly in the sense that the inner cycle of checking the inequality* (9) *is finite.*

*Proof.* Since, before each check of the inequality (9) on the step $k$, we multiply $M_k$ by 2, after finite number of these multiplications, we will have $M_k \geq L$. Since $\varphi$ has $L$-Lipschitz-continuous gradient, due to (3), we obtain that (9) holds after finite number of these repetitions. $\qquad\square$

**Lemma 2.** *Let the sequences* $\{\lambda_k, \eta_k, \zeta_k, \alpha_k, \beta_k\}$, $k \geq 0$ *be generated by Algorithm 1. Then, for all* $\lambda \in \Lambda$, *it holds that*

$$\alpha_{k+1}\langle \nabla\varphi(\lambda_{k+1}), \zeta_k - \lambda \rangle \leq \beta_{k+1}(\varphi(\lambda_{k+1}) - \varphi(\eta_{k+1})) + V[\zeta_k](\lambda) - V[\zeta_{k+1}](\lambda). \tag{10}$$

*Proof.* Note that, from the optimality condition in (7), for any $\lambda \in \Lambda$, we have

$$\langle \nabla V[\zeta_k](\zeta_{k+1}) + \alpha_{k+1}\nabla\varphi(\lambda_{k+1}), \lambda - \zeta_{k+1}\rangle \geq 0. \tag{11}$$

By the definition of $V[\zeta](\lambda)$, we obtain, for any $\lambda \in \Lambda$,

$$\begin{aligned}
V[\zeta_k](\lambda) - V[\zeta_{k+1}](\lambda) - V[\zeta_k](\zeta_{k+1}) =&\, d(\lambda) - d(\zeta_k) - \langle \nabla d(\zeta_k), \lambda - \zeta_k \rangle \\
&- (d(\lambda) - d(\zeta_{k+1}) - \langle \nabla d(\zeta_{k+1}), \lambda - \zeta_{k+1} \rangle) \\
&- (d(\zeta_{k+1}) - d(\zeta_k) - \langle \nabla d(\zeta_k), \zeta_{k+1} - \zeta_k \rangle) \\
=&\, \langle \nabla d(\zeta_k) - \nabla d(\zeta_{k+1}), \zeta_{k+1} - \lambda \rangle \\
=&\, \langle -\nabla V[\zeta_k](\zeta_{k+1}), \zeta_{k+1} - \lambda \rangle. 
\end{aligned} \tag{12}$$

**Algorithm 1** Adaptive Accelerated Gradient Descent (AAGD)

---

**Input:** starting point $\lambda_0 \in \Lambda^0$, initial guess $0 < L_0 < 2L$, prox-setup: $d(\lambda)$ – 1-strongly convex w.r.t. $\|\cdot\|$, $V[\zeta](\lambda) := d(\lambda) - d(\zeta) - \langle \nabla d(\zeta), \lambda - \zeta \rangle$, $\lambda \in \Lambda, \zeta \in \Lambda^0$.

1: Set $k = 0$, $\beta_0 = \alpha_0 = 0$, $\eta_0 = \zeta_0 = \lambda_0$.
2: **repeat**
3:    Set $M_k = L_k/2$.
4:    **repeat**
5:      Set $M_k = 2M_k$, find $\alpha_{k+1}$ as the largest root of the equation

$$\beta_{k+1} := \beta_k + \alpha_{k+1} = M_k \alpha_{k+1}^2. \tag{5}$$

6:

$$\lambda_{k+1} = \frac{\alpha_{k+1}\zeta_k + \beta_k \eta_k}{\beta_{k+1}}. \tag{6}$$

7:

$$\zeta_{k+1} = \arg\min_{\lambda \in \Lambda} \{V[\zeta_k](\lambda) + \alpha_{k+1}(\varphi(\lambda_{k+1}) + \langle \nabla\varphi(\lambda_{k+1}), \lambda - \lambda_{k+1} \rangle)\}. \tag{7}$$

8:

$$\eta_{k+1} = \frac{\alpha_{k+1}\zeta_{k+1} + \beta_k \eta_k}{\beta_{k+1}}. \tag{8}$$

9:    **until**

$$\varphi(\eta_{k+1}) \leq \varphi(\lambda_{k+1}) + \langle \nabla\varphi(\lambda_{k+1}), \eta_{k+1} - \lambda_{k+1} \rangle + \frac{M_k}{2}\|\eta_{k+1} - \lambda_{k+1}\|^2. \tag{9}$$

10:   Set $L_{k+1} = M_k/2$, $k = k + 1$.
11: **until** Option 1: $k = k_{max}$.
     Option 2: $R^2/\beta_k \leq \varepsilon$.
     Option 3:

$$\varphi(\eta_k) - \min_{\lambda \in \Lambda : V[\zeta_0](\lambda) \leq R^2} \left\{ \sum_{i=0}^{k} \frac{\alpha_i}{\beta_k} \left( \varphi(\lambda_i) + \langle \nabla\varphi(\lambda_i), \lambda - \lambda_i \rangle \right) \right\} \leq \varepsilon.$$

Here $R$ is such that $V[\zeta_0](\lambda_*) \leq R^2$ and $\varepsilon$ is the desired accuracy.

**Output:** The point $\eta_{k+1}$.

---

Further, for any $\lambda \in \Lambda$,

$$\alpha_{k+1}\langle \nabla\varphi(\lambda_{k+1}), \zeta_k - \lambda \rangle = \alpha_{k+1}\langle \nabla\varphi(\lambda_{k+1}), \zeta_k - \zeta_{k+1} \rangle + \alpha_{k+1}\langle \nabla\varphi(\lambda_{k+1}), \zeta_{k+1} - \lambda \rangle$$

$$\overset{(11)}{\leq} \alpha_{k+1}\langle \nabla\varphi(\lambda_{k+1}), \zeta_k - \zeta_{k+1} \rangle + \langle -\nabla V[\zeta_k](\zeta_{k+1}), \zeta_{k+1} - \lambda \rangle$$

$$\overset{(12)}{=} \alpha_{k+1}\langle \nabla\varphi(\lambda_{k+1}), \zeta_k - \zeta_{k+1} \rangle + V[\zeta_k](\lambda) - V[\zeta_{k+1}](\lambda) - V[\zeta_k](\zeta_{k+1})$$

$$\overset{(4)}{\leq} \alpha_{k+1}\langle \nabla\varphi(\lambda_{k+1}), \zeta_k - \zeta_{k+1} \rangle + V[\zeta_k](\lambda) - V[\zeta_{k+1}](\lambda) - \frac{1}{2}\|\zeta_k - \zeta_{k+1}\|^2$$

$$\overset{(6),(8)}{=} \beta_{k+1}\langle \nabla\varphi(\lambda_{k+1}), \lambda_{k+1} - \eta_{k+1} \rangle + V[\zeta_k](\lambda) - V[\zeta_{k+1}](\lambda) - \frac{\beta_{k+1}^2}{2\alpha_{k+1}^2}\|\lambda_{k+1} - \eta_{k+1}\|^2$$

$$\overset{(5)}{=} \beta_{k+1}\left( \langle \nabla\varphi(\lambda_{k+1}), \lambda_{k+1} - \eta_{k+1} \rangle - \frac{M_k}{2}\|\lambda_{k+1} - \eta_{k+1}\|^2 \right) + V[\zeta_k](\lambda) - V[\zeta_{k+1}](\lambda)$$

$$\overset{(9)}{\leq} \beta_{k+1}\left( \varphi(\lambda_{k+1}) - \varphi(\eta_{k+1}) \right) + V[\zeta_k](\lambda) - V[\zeta_{k+1}](\lambda).$$

$\square$

**Lemma 3.** *Let the sequences* $\{\lambda_k, \eta_k, \zeta_k, \alpha_k, \beta_k\}$, $k \geq 0$ *be generated by Algorithm 1. Then, for all* $\lambda \in \Lambda$, *it holds that*

$$\beta_{k+1}\varphi(\eta_{k+1}) - \beta_k\varphi(\eta_k) \leq \alpha_{k+1}\left(\varphi(\lambda_{k+1}) + \langle\nabla\varphi(\lambda_{k+1}), \lambda - \lambda_{k+1}\rangle\right) + V[\zeta_k](\lambda) - V[\zeta_{k+1}](\lambda). \tag{13}$$

*Proof.* For any $\lambda \in \Lambda$,

$$
\begin{aligned}
\alpha_{k+1}\langle\nabla\varphi(\lambda_{k+1}), \lambda_{k+1} - \lambda\rangle &= \alpha_{k+1}\langle\nabla\varphi(\lambda_{k+1}), \lambda_{k+1} - \zeta_k\rangle + \alpha_{k+1}\langle\nabla\varphi(\lambda_{k+1}), \zeta_k - \lambda\rangle \\
&\overset{(5),(6)}{=} \beta_k\langle\nabla\varphi(\lambda_{k+1}), \eta_k - \lambda_{k+1}\rangle + \alpha_{k+1}\langle\nabla\varphi(\lambda_{k+1}), \zeta_k - \lambda\rangle \\
&\overset{\text{conv-ty}}{\leq} \beta_k\left(\varphi(\eta_k) - \varphi(\lambda_{k+1})\right) + \alpha_{k+1}\langle\nabla\varphi(\lambda_{k+1}), \zeta_k - \lambda\rangle \\
&\overset{(10)}{\leq} \beta_k\left(\varphi(\eta_k) - \varphi(\lambda_{k+1})\right) + \beta_{k+1}\left(\varphi(\lambda_{k+1}) - \varphi(\eta_{k+1})\right) + V[\zeta_k](\lambda) - V[\zeta_{k+1}](\lambda) \\
&= \alpha_{k+1}\varphi(\lambda_{k+1}) + \beta_k\varphi(\eta_k) - \beta_{k+1}\varphi(\eta_{k+1}) + V[\zeta_k](\lambda) - V[\zeta_{k+1}](\lambda). \tag{14}
\end{aligned}
$$

Rearranging terms, we obtain the statement of the Lemma. $\square$

**Theorem 1.** *Let the sequences* $\{\lambda_k, \eta_k, \zeta_k, \alpha_k, \beta_k\}$, $k \geq 0$ *be generated by Algorithm 1. Then, for all* $k \geq 0$, *it holds that*

$$\beta_k\varphi(\eta_k) \leq \min_{\lambda \in \Lambda}\left\{\sum_{i=0}^{k}\alpha_i\left(\varphi(\lambda_i) + \langle\nabla\varphi(\lambda_i), \lambda - \lambda_i\rangle\right) + V[\zeta_0](\lambda)\right\}. \tag{15}$$

*The number of inner cycle iterations after an iteration* $k \geq 0$ *does not exceed*

$$4k + 4 + 2\log_2\left(\frac{L}{L_0}\right), \tag{16}$$

*where $L$ is the Lipschitz constant for the gradient of* $\varphi$.

*Proof.* Let us change the counter in Lemma 2 from $k$ to $i$ and sum all the inequalities for $i = 0, ..., k-1$. Then, for any $\lambda \in \Lambda$,

$$\beta_k\varphi(\eta_k) - \beta_0\varphi(\eta_0) \leq \sum_{i=0}^{k-1}\alpha_{i+1}\left(\varphi(\lambda_{i+1}) + \langle\nabla\varphi(\lambda_{i+1}), \lambda - \lambda_{i+1}\rangle\right) + V[\zeta_0](\lambda) - V[\zeta_k](\lambda). \tag{17}$$

Whence, since $\beta_0 = \alpha_0 = 0$ and $V[\zeta_k](\lambda) \geq 0$,

$$\beta_k\varphi(\eta_k) \leq \sum_{i=0}^{k}\alpha_i\left(\varphi(\lambda_i) + \langle\nabla\varphi(\lambda_i), \lambda - \lambda_i\rangle\right) + V[\zeta_0](\lambda), \quad \lambda \in \Lambda. \tag{18}$$

Taking in the right hand side the minimum in $\lambda \in \Lambda$, we obtain the first statement of the Theorem.

The second statement of the Theorem is proved in the same way as in Nesterov and Polyak [2006], but we provide the proof for the reader's convenience. Let us again change the iteration counter in Algorithm 1 from $k$ to $i$. Let $j_i \geq 1$ be the total number of checks of the inequality (9) on the step $i \geq 0$. Then, $j_0 = 1 + \log_2 \frac{M_0}{L_0}$ and, for $i \geq 1$, $M_i = 2^{j_i - 1} L_i = 2^{j_i - 1} \frac{M_{i-1}}{2}$. Thus, $j_i = 2 + \log_2 \frac{M_i}{M_{i-1}}$, $i \geq 1$. Further, by the same reasoning as in Lemma 2, we obtain that $M_i \leq 2L$, $i \geq 0$. Then, the total number of checks of the inequality (9) is

$$\sum_{i=0}^{k} j_i = 1 + \log_2 \frac{M_0}{L_0} + \sum_{i=1}^{k} \left( 2 + \log_2 \frac{M_i}{M_{i-1}} \right) = 2k + 1 + \log_2 \frac{M_k}{L_0} \leq 2k + 2 + \log_2 \frac{L}{L_0}.$$

At the same time, each check of the inequality (9) requires two oracle calls. This proves the second statement of the Theorem. $\qquad \square$

**Corollary 1.** *Let the sequences $\{\lambda_k, \eta_k, \zeta_k, \alpha_k, \beta_k\}$, $k \geq 0$ be generated by Algorithm 1. Then, for all $k \geq 0$, it holds that*

$$\varphi(\eta_k) - \min_{\lambda \in \Lambda} \varphi(\lambda) \leq \frac{V[\zeta_0](\lambda_*)}{\beta_k}, \tag{19}$$

*where $\lambda_*$ is the solution of $\min_{\lambda \in \Lambda} \varphi(\lambda)$ s.t. $V[\zeta_0](\lambda_*)$ is minimal among all the solutions.*

*Proof.* Let $\lambda_*$ be the solution of $\min_{\lambda \in \Lambda} \varphi(\lambda)$ s.t. $V[\zeta_0](\lambda_*)$ is minimal among all the solutions. Using convexity of $\varphi$, from Theorem 1, we obtain

$$\beta_k \varphi(\eta_k) \leq \sum_{i=0}^{k} \alpha_i \varphi(\lambda_*) + V[\zeta_0](\lambda_*).$$

Since $\beta_k = \sum_{i=0}^{k} \alpha_i$, we obtain the statement of the Corollary. $\qquad \square$

The following Corollary justifies the stopping criteria in Algorithm 1.

**Corollary 2.** *Let $\lambda_*$ be a solution of $\min_{\lambda \in \Lambda} \varphi(\lambda)$ such that $V[\zeta_0](\lambda_*)$ is minimal among all the solutions. Let $R$ be such that $V[\zeta_0](\lambda_*) \leq R^2$ and $\varepsilon$ be the desired accuracy. Let the sequences $\{\lambda_k, \eta_k, \zeta_k, \alpha_k, \beta_k\}$, $k \geq 0$ be generated by Algorithm 1. Then, if one of the following inequalities holds*

$$R^2 / \beta_k \leq \varepsilon, \tag{20}$$

$$\varphi(\eta_k) - \min_{\lambda \in \Lambda : V[\zeta_0](\lambda) \leq R^2} \left\{ \sum_{i=0}^{k} \frac{\alpha_i}{\beta_k} \left( \varphi(\lambda_i) + \langle \nabla \varphi(\lambda_i), \lambda - \lambda_i \rangle \right) \right\} \leq \varepsilon, \tag{21}$$

*then*

$$\varphi(\eta_k) - \min_{\lambda \in \Lambda} \varphi(\lambda) \leq \varepsilon. \tag{22}$$

*Proof.* If the inequality (20) holds, the statement of the Corollary follows from inequality $V[\zeta_0](\lambda_*) \leq R^2$ Corollary 1.

Since $V[\zeta_0](\lambda_*) \leq R^2$, the point $\lambda_*$ is a feasible point in the problem

$$\min_{\lambda \in \Lambda : V[\zeta_0](\lambda) \leq R^2} \left\{ \sum_{i=0}^{k} \frac{\alpha_i}{\beta_k} \left( \varphi(\lambda_i) + \langle \nabla \varphi(\lambda_i), \lambda - \lambda_i \rangle \right) \right\}.$$

Then, by convexity of $\varphi$, we obtain

$$\min_{\lambda \in \Lambda : V[\zeta_0](\lambda) \leq R^2} \left\{ \sum_{i=0}^{k} \frac{\alpha_i}{\beta_k} \left( \varphi(\lambda_i) + \langle \nabla \varphi(\lambda_i), \lambda - \lambda_i \rangle \right) \right\} \leq \sum_{i=0}^{k} \frac{\alpha_i}{\beta_k} \left( \varphi(\lambda_i) + \langle \nabla \varphi(\lambda_i), \lambda_* - \lambda_i \rangle \right)$$
$$\leq \varphi(\lambda_*).$$

This and (21) finishes the proof. $\qquad \square$

Let us now obtain the lower bound for the sequence $\beta_k$, $k \geq 0$, which will give the rate of convergence for Algorithm 1.

**Lemma 4.** *Let the sequence* $\{\beta_k\}$, *$k \geq 0$ be generated by Algorithm 1. Then, for all $k \geq 1$ it holds that*

$$\beta_k \geq \frac{(k+1)^2}{8L}, \tag{23}$$

*where $L$ is the Lipschitz constant for the gradient of $\varphi$.*

*Proof.* As we mentioned in the proof of Theorem 1, $M_k \leq 2L$, $k \geq 0$. For $k = 1$, since $\alpha_0 = 0$ and $A_1 = \alpha_0 + \alpha_1 = \alpha_1$, we have from (5)

$$\beta_1 = \alpha_1 = \frac{1}{M_1} \geq \frac{1}{2L}.$$

Hence, (23) holds for $k = 1$.

Let us now assume that (23) holds for some $k \geq 1$ and prove that it holds for $k+1$. From (5) we have a quadratic equation for $\alpha_{k+1}$

$$M_k \alpha_{k+1}^2 - \alpha_{k+1} - \beta_k = 0.$$

Since we need to take the largest root, we obtain,

$$\alpha_{k+1} = \frac{1 + \sqrt{1 + 4M_k \beta_k}}{2M_k} = \frac{1}{2M_k} + \sqrt{\frac{1}{4M_k^2} + \frac{\beta_k}{M_k}} \geq \frac{1}{2M_k} + \sqrt{\frac{\beta_k}{M_k}}$$
$$\geq \frac{1}{4L} + \frac{1}{\sqrt{2L}} \frac{k+1}{2\sqrt{2L}} = \frac{k+2}{4L},$$

where we used the induction assumption that (23) holds for $k$. Using the obtained inequality, from (5) and (23) for $k$, we get

$$\beta_{k+1} = \beta_k + \alpha_{k+1} \geq \frac{(k+1)^2}{8L} + \frac{k+2}{4L} \geq \frac{(k+2)^2}{8L}.$$

$\qquad \square$

**Corollary 3.** *Let the sequences $\{\lambda_k, \eta_k, \zeta_k\}$, $k \geq 0$ be generated by Algorithm 1. Then, for all $k \geq 1$, it holds that*

$$\varphi(\eta_k) - \min_{\lambda \in \Lambda} \varphi(\lambda) \leq \frac{8LV[\zeta_0](\lambda_*)}{(k+1)^2}, \tag{24}$$

*where $\lambda_*$ is the solution of $\min_{\lambda \in \Lambda} \varphi(\lambda)$ s.t. $V[\zeta_0](\lambda_*)$ is minimal among all the solutions.*

## 2.4 Adaptive Primal-Dual Accelerated Gradient Descent for Constrained Convex Optimization

In this section, we return to the constrained convex optimization problem, which was considered in Section 3 of the main part of the paper. For the reader's convenience, we repeat the problem statement and some details.

### 2.4.1 Preliminaries

We consider convex optimization problem of the following form

$$(P_1) \qquad \min_{x \in Q \subseteq E} \{f(x) : Ax = b\},$$

where $f(x)$ is a $\gamma$-strongly convex function on $Q$ with respect to some chosen norm $\|\cdot\|_E$ on $E$ and $A : E \to H$ is a linear operator, $b \in H$.

The Lagrange dual problem to Problem $(P_1)$ is

$$(D_1) \qquad \max_{\lambda \in \Lambda} \left\{ -\langle \lambda, b \rangle + \min_{x \in Q} \left( f(x) + \langle A^T \lambda, x \rangle \right) \right\}.$$

Here we denote $\Lambda = H^*$ the space of Lagrange multipliers. It is convenient to rewrite Problem $(D_1)$ in the equivalent form of a minimization problem

$$(P_2) \qquad \min_{\lambda \in \Lambda} \left\{ \langle \lambda, b \rangle + \max_{x \in Q} \left( -f(x) - \langle A^T \lambda, x \rangle \right) \right\}.$$

It is obvious that

$$Opt[D_1] = -Opt[P_2], \tag{25}$$

where $Opt[D_1]$, $Opt[P_2]$ are the optimal function value in Problem $(D_1)$ and Problem $(P_2)$ respectively. The following inequality follows from the weak duality

$$Opt[P_1] \geq Opt[D_1]. \tag{26}$$

We denote

$$\varphi(\lambda) = \langle \lambda, b \rangle + \max_{x \in Q} \left( -f(x) - \langle A^T \lambda, x \rangle \right). \tag{27}$$

Since $f$ is strongly convex, $\varphi(\lambda)$ is a smooth function and its gradient is equal to (see e.g. Nesterov [2005])

$$\nabla \varphi(\lambda) = b - Ax(\lambda), \tag{28}$$

where $x(\lambda)$ is the unique solution of the strongly-convex problem

$$\max_{x \in Q} \left( -f(x) - \langle A^T \lambda, x \rangle \right). \tag{29}$$

Note that $\nabla \varphi(\lambda)$ is Lipschitz-continuous (see e.g. Nesterov [2005]) with constant

$$L \leq \frac{\|A\|_{E \to H}^2}{\gamma}.$$

We also assume that the dual problem $(D_1)$ has a solution $\lambda^*$ and there exists some $R > 0$ such that

$$\|\lambda^*\|_2 \leq R < +\infty. \tag{30}$$

### 2.4.2 Adaptive Primal-Dual Accelerated Gradient Descent

Now we are ready to apply Algorithm 1 to the problem $(P_2)$ and incorporate in the algorithm a procedure, which allows to reconstruct also an approximate solution of the problem $(P_1)$. We choose Euclidean proximal setup, which means that we introduce euclidean norm $\|\cdot\|_2$ in the space of vectors $\lambda$ and choose the prox-function $d(\lambda) = \frac{1}{2}\|\lambda\|_2^2$. Then, we have $V[\zeta](\lambda) = \frac{1}{2}\|\lambda - \zeta\|_2^2$. We state here as Algorithm 2 a more detailed version of Algorithm 3 in the main part of the paper. The first difference is that here we do not introduce an auxiliary sequence $\tau_k = \alpha_{k+1}/\beta_{k+1}$. The second difference is that here we use an equivalent form

$$\zeta_{k+1} = \arg\min_{\lambda \in \Lambda} \left\{ \frac{1}{2}\|\lambda - \zeta_k\|_2^2 + \alpha_{k+1}(\varphi(\lambda_{k+1}) + \langle \nabla \varphi(\lambda_{k+1}), \lambda - \lambda_{k+1} \rangle) \right\}$$

of the step

$$\zeta_{k+1} = \zeta_k - \alpha_{k+1} \nabla \varphi(\lambda_{k+1}).$$

The third difference consists in the observation that, since, by definition, $\beta_k = \sum_{i=0}^{k} \alpha_k$,

$$\hat{x}_{k+1} = \frac{\alpha_{k+1} x(\lambda_{k+1}) + \beta_k \hat{x}_k}{\beta_{k+1}} = \frac{1}{\beta_{k+1}} \sum_{i=0}^{k+1} \alpha_i x(\lambda_i).$$

Finally, here we use a stronger stopping rule

$$|f(\hat{x}_{k+1}) + \varphi(\eta_{k+1})| \leq \varepsilon_f$$

than in the main part of the paper. The reason is that, to obtain complexity result for approximating OT distance, it is enough to satisfy

$$f(\hat{x}_{k+1}) + \varphi(\eta_{k+1}) \leq \varepsilon$$

with a special choice of $\varepsilon$.

**Algorithm 2** Adaptive Primal-Dual Accelerated Gradient Descent (APDAGD)

---

**Input:** starting point $\lambda_0 = 0$, initial guess $L_0 > 0$, accuracy $\varepsilon_f, \varepsilon_{eq} > 0$.

1: Set $k = 0$, $\beta_0 = \alpha_0 = 0$, $\eta_0 = \zeta_0 = \lambda_0 = 0$.
2: **repeat**
3:     Set $M_k = L_k/2$.
4:     **repeat**
5:         Set $M_k = 2M_k$, find $\alpha_{k+1}$ as the largest root of the equation

$$\beta_{k+1} := \beta_k + \alpha_{k+1} = M_k \alpha_{k+1}^2. \tag{31}$$

6:         Calculate

$$\lambda_{k+1} = \frac{\alpha_{k+1}\zeta_k + \beta_k \eta_k}{\beta_{k+1}}. \tag{32}$$

7:         Calculate

$$\zeta_{k+1} = \arg\min_{\lambda \in \Lambda} \left\{ \frac{1}{2}\|\lambda - \zeta_k\|_2^2 + \alpha_{k+1}(\varphi(\lambda_{k+1}) + \langle \nabla\varphi(\lambda_{k+1}), \lambda - \lambda_{k+1}\rangle) \right\}. \tag{33}$$

8:         Calculate

$$\eta_{k+1} = \frac{\alpha_{k+1}\zeta_{k+1} + \beta_k \eta_k}{\beta_{k+1}}. \tag{34}$$

9:     **until**

$$\varphi(\eta_{k+1}) \leq \varphi(\lambda_{k+1}) + \langle \nabla\varphi(\lambda_{k+1}), \eta_{k+1} - \lambda_{k+1}\rangle + \frac{M_k}{2}\|\eta_{k+1} - \lambda_{k+1}\|_2^2. \tag{35}$$

10:    Set

$$\hat{x}_{k+1} = \frac{1}{\beta_{k+1}}\sum_{i=0}^{k+1}\alpha_i x(\lambda_i) = \frac{\alpha_{k+1}x(\lambda_{k+1}) + \beta_k \hat{x}_k}{\beta_{k+1}}.$$

11:    Set $L_{k+1} = M_k/2$, $k = k + 1$.
12: **until** $|f(\hat{x}_{k+1}) + \varphi(\eta_{k+1})| \leq \varepsilon_f$, $\|A_1\hat{x}_{k+1} - b_1\|_2 \leq \varepsilon_{eq}$.

**Output:** The points $\hat{x}_{k+1}$, $\eta_{k+1}$.

---

**Theorem 2.** *Assume that the objective in the problem $(P_1)$ is $\gamma$-strongly convex and that the dual solution $\lambda^*$ satisfies $\|\lambda^*\|_2 \leq R$. Then, for $k \geq 1$, the points $\hat{x}_k$, $\eta_k$ in Algorithm 2 satisfy*

$$-\frac{16LR^2}{k^2} \leq f(\hat{x}_k) - Opt[P_1] \leq f(\hat{x}_k) + \varphi(\eta_k) \leq \frac{16LR^2}{k^2}, \tag{36}$$

$$\|A\hat{x}_k - b\|_2 \leq \frac{16LR}{k^2}, \tag{37}$$

$$\|\hat{x}_k - x^*\|_E \leq \frac{8}{k}\sqrt{\frac{LR^2}{\gamma}}, \tag{38}$$

*where $x^*$ and $Opt[P_1]$ are respectively an optimal solution and the optimal value in the problem $(P_1)$, and $L \leq \frac{\|A\|_{E\to H}^2}{\gamma}$. Moreover, the stopping criterion in step 11 is correctly defined and* *the number of inner cycle iterations after an iteration $k \geq 0$ does not exceed*

$$4k + 4 + 2\log_2\left(\frac{L}{L_0}\right),\tag{39}$$

*where $L \leq \frac{\|A\|_{E \to H}^2}{\gamma}$ is the Lipschitz constant for the gradient of $\varphi$.*

*Proof.* From Theorem 1 with specific choice of the Bregman divergence, since $\zeta_0 = 0$, we have, for all $k \geq 0$,

$$\beta_k \varphi(\eta_k) \leq \min_{\lambda \in \Lambda} \left\{ \sum_{i=0}^{k} \alpha_i \left( \varphi(\lambda_i) + \langle \nabla \varphi(\lambda_i), \lambda - \lambda_i \rangle \right) + \frac{1}{2}\|\lambda\|_2^2 \right\}\tag{40}$$

Let us introduce a set $\Lambda_R = \{\lambda : \|\lambda\|_2 \leq 2R\}$ where $R$ is given in (30). Then, from (40), we obtain

$$\begin{aligned}
\beta_k \varphi(\eta_k) &\leq \min_{\lambda \in \Lambda} \left\{ \sum_{i=0}^{k} \alpha_i \left( \varphi(\lambda_i) + \langle \nabla \varphi(\lambda_i), \lambda - \lambda_i \rangle \right) + \frac{1}{2}\|\lambda\|_2^2 \right\} \\
&\leq \min_{\lambda \in \Lambda_R} \left\{ \sum_{i=0}^{k} \alpha_i \left( \varphi(\lambda_i) + \langle \nabla \varphi(\lambda_i), \lambda - \lambda_i \rangle \right) + \frac{1}{2}\|\lambda\|_2^2 \right\} \\
&\leq \min_{\lambda \in \Lambda_R} \left\{ \sum_{i=0}^{k} \alpha_i \left( \varphi(\lambda_i) + \langle \nabla \varphi(\lambda_i), \lambda - \lambda_i \rangle \right) \right\} + 2R^2.
\end{aligned}\tag{41}$$

On the other hand, from the definition (27) of $\varphi(\lambda)$, we have

$$\begin{aligned}
\varphi(\lambda_i) &= \langle \lambda_i, b \rangle + \max_{x \in Q} \left( -f(x) - \langle A^T \lambda_i, x \rangle \right) \\
&= \langle \lambda_i, b \rangle - f(x(\lambda_i)) - \langle A^T \lambda_i, x(\lambda_i) \rangle.
\end{aligned}$$

Combining this equality with (28), we obtain

$$\begin{aligned}
\varphi(\lambda_i) - \langle \nabla \varphi(\lambda_i), \lambda_i \rangle &= \varphi(\lambda) - \langle \nabla \varphi(\lambda), \lambda_i \rangle \\
&= \langle \lambda_i, b \rangle - f(x(\lambda_i)) - \langle A^T \lambda_i, x(\lambda_i) \rangle \\
&\quad - \langle b - Ax(\lambda_i), \lambda_i \rangle = -f(x(\lambda_i)).
\end{aligned}$$

Summing these inequalities from $i = 0$ to $i = k$ with the weights $\{\alpha_i\}_{i=1,\ldots k}$, we get, using the convexity of $f$,

$$\begin{aligned}
&\sum_{i=0}^{k} \alpha_i \left( \varphi(\lambda_i) + \langle \nabla \varphi(\lambda_i), \lambda - \lambda_i \rangle \right) \\
&= -\sum_{i=0}^{k} \alpha_i f(x(\lambda_i)) + \sum_{i=0}^{k} \alpha_i \langle b - Ax(\lambda_i), \lambda \rangle \\
&\leq -\beta_k f(\hat{x}_k) + \beta_k \langle b - A\hat{x}_k, \lambda \rangle.
\end{aligned}$$

Substituting this inequality to (41), we obtain

$$\beta_k \varphi(\eta_k) \leq -\beta_k f(\hat{x}_k) + \beta_k \min_{\lambda \in \Lambda_R} \{\langle b - A\hat{x}_k, \lambda \rangle\} + 2R^2.$$

Finally, since

$$\max_{\lambda \in \Lambda_R} \{\langle -b + A_1 \hat{x}_k, \lambda \rangle\} = 2R \|A\hat{x}_k - b\|_2,$$

we obtain

$$\varphi(\eta_k) + f(\hat{x}_k) + 2R \|A\hat{x}_k - b\|_2 \leq \frac{2R^2}{\beta_k}. \tag{42}$$

Since $\lambda^*$ is an optimal solution of Problem $(D_1)$, we have, for any $x \in Q$

$$Opt[P_1] \leq f(x) + \langle \lambda^*, Ax - b \rangle.$$

Using the assumption (30), we get

$$f(\hat{x}_k) \geq Opt[P_1] - R \|A\hat{x}_k - b\|_2. \tag{43}$$

Hence,

$$\begin{aligned}
\varphi(\eta_k) + f(\hat{x}_k) &= \varphi(\eta_k) - Opt[P_2] + Opt[P_2] + Opt[P_1] - Opt[P_1] + f(\hat{x}_k) \\
&\overset{(25)}{=} \varphi(\eta_k) - Opt[P_2] - Opt[D_1] + Opt[P_1] - Opt[P_1] + f(\hat{x}_k) \\
&\overset{(26)}{\geq} -Opt[P_1] + f(\hat{x}_k) \overset{(43)}{\geq} -R \|A\hat{x}_k - b\|_2.
\end{aligned} \tag{44}$$

This and (42) give

$$R \|A\hat{x}_k - b\|_2 \leq \frac{2R^2}{\beta_k}. \tag{45}$$

Hence, we obtain

$$\varphi(\eta_k) + f(\hat{x}_k) \overset{(44),(45)}{\geq} -\frac{2R^2}{\beta_k}. \tag{46}$$

On the other hand, we have

$$\varphi(\eta_k) + f(\hat{x}_k) \overset{(42)}{\leq} \frac{2R^2}{\beta_k}. \tag{47}$$

Combining (45), (46), (47), we conclude

$$\begin{aligned}
\|A\hat{x}_k - b\|_2 &\leq \frac{2R}{\beta_k}, \\
|\varphi(\eta_k) + f(\hat{x}_k)| &\leq \frac{2R^2}{\beta_k}.
\end{aligned} \tag{48}$$

At the same time,

$$\varphi(\eta_k) + Opt[P_1] = \varphi(\eta_k) - Opt[P_2] + Opt[P_2] + Opt[P_1]$$
$$\overset{(25)}{=} \varphi(\eta_k) - Opt[P_2] - Opt[D_1] + Opt[P_1] \overset{(26)}{\geq} 0.$$

Hence,

$$f(\hat{x}_k) - Opt[P_1] \leq f(\hat{x}_k) + \varphi(\eta_k). \qquad (49)$$

From (48), (49), by Lemma 4, stating that, for any $k \geq 0$, $\beta_k \geq \frac{(k+1)^2}{8L}$, we obtain inequalities (36) and (37) in the Theorem statements.

It remains to prove inequality (38). By the optimality condition for Problem $(P_1)$, we have

$$\langle \nabla f(x^*) + A^T \lambda^*, \hat{x}_k - x^* \rangle \geq 0, \quad Ax^* = b,$$

where $\nabla f(x^*) \in \partial f(x^*)$. Then

$$\langle \nabla f(x^*), \hat{x}_k - x^* \rangle \geq -\langle A^T \lambda^*, \hat{x}_k - x^* \rangle$$
$$\geq -\langle \lambda^{*(1)}, A\hat{x}_k - b \rangle$$
$$\geq -R\|A\hat{x}_k - b\|_2 \overset{(45)}{\geq} -\frac{2R^2}{\beta_k}, \qquad (50)$$

where we used the same reasoning as while deriving (43). Using this inequality and $\gamma$ strong convexity of $f$, we obtain

$$\frac{\gamma}{2}\|\hat{x}_k - x^*\|_E^2 \leq f(\hat{x}_k) - Opt[P_1] - \langle \nabla f(x^*), \hat{x}_k - x^* \rangle \overset{(48),(49)}{\leq} \frac{4R^2}{\beta_k}.$$

Since, by Lemma 4, for any $k \geq 0$, $\beta_k \geq \frac{(k+1)^2}{8L}$, we obtain inequality (38).

$\square$

# References

Aaron Ben-Tal and Arkadi Nemirovski. *Lectures on Modern Convex Optimization (Lecture Notes)*. Personal web-page of A. Nemirovski, 2015. URL `http://www2.isye.gatech.edu/~nemirovs/Lect_ModConvOpt.pdf`.

Guanghui Lan, Zhaosong Lu, and Renato D. C. Monteiro. Primal-dual first-order methods with $O(1/\varepsilon)$ iteration-complexity for cone programming. *Mathematical Programming*, 126 (1):1–29, 2011.

Yurii Nesterov. Smooth minimization of non-smooth functions. *Mathematical Programming*, 103(1):127–152, 2005.

Yurii Nesterov and Boris Polyak. Cubic regularization of newton method and its global performance. *Mathematical Programming*, 108(1):177–205, 2006. ISSN 1436-4646. doi: 10.1007/s10107-006-0706-8. URL `http://dx.doi.org/10.1007/s10107-006-0706-8`.

Paul Tseng. On accelerated proximal gradient methods for convex-concave optimization. Technical report, MIT, 2008. URL `http://www.mit.edu/~dimitrib/PTseng/papers/apgm.pdf`.